\documentstyle[12pt]{article}

\newcommand{\hephypreprint}[4]
{\noindent\begin{minipage}[t]{\textwidth}\begin{center}
\framebox[\textwidth]
{$\rule[6mm]{0mm}{0mm}$ \raisebox{1.3mm}
{#1{Institut f\"ur Hochenergiephysik der \"Osterreichischen Akademie
der Wissenschaften}}
}

\vspace{2mm}    \rule{\textwidth}{0.2mm}\\
\vspace{-4mm}   \rule{\textwidth}{1pt}
\mbox{ }    #2    \hfill    #3   \mbox{ }\\
\vspace{-2mm}   \rule{\textwidth}{1pt}\\
\vspace{-4.2mm} \rule{\textwidth}{0.2mm}
\end{center}
\end{minipage}

\vspace{20mm}
\begin{center} {#4}  \end{center}
}

\begin{document}

\pagestyle{empty}
\setcounter{page}{0}
\hephypreprint{\small\bf}{February 1995}{HEPHY--PUB 618/95, AZPH--TH/95-03}
{\large\bf
         Wavelet correlations in hierarchical branching processes
}

\begin{center}
\vspace{1.5cm}
Martin Greiner\footnote{email address: tp21@ddagsi3.gsi.de}
and
Jens Giesemann\footnote{email address:
                        jens.giesemann@physik.uni-giessen.de}, \\
{\em Institut f\"ur Theoretische Physik, Justus Liebig Universit\"at,
35392 Giessen, Germany,} \\
\vspace{0.5cm}
Peter Lipa\footnote{email address: lipa@hephy.oeaw.ac.at}, \\
{\em Institut f\"ur Hochenergiephysik der \"Osterreichischen Akademie
der Wissenschaften, \\
Nikolsdorfergasse 18, A-1050 Wien, Austria,} \\
\vspace{0.5cm}
Peter Carruthers\footnote{email address: carruthers@ccit.arizona.edu},
\\
{\em Department of Physics, University of Arizona, Tucson AZ 85721,
USA.} \\
\end{center}

\vspace{2cm}

\begin{quote}
\centerline{ABSTRACT}
A study of correlations in tractable multiparticle cascade models in
terms of
wavelets reveals many promising features. The selfsimilar construction
of the wavelet basis functions and their multiscale localization
properties provide a new approach to the statistical analysis and
analytical control of hierarchically organized branching processes. The
exact analytical solution of several discrete models shows that the
wavelet
transformation supresses redundancy in the correlation information.
Wavelet correlations can be naturally interpreted as correlations
between structures (clumps) living on different scales.
\end{quote}

\vfill\noindent
\mbox{} \hfill (submitted to Z.Phys.C)

\newpage

\pagestyle{plain}
\section{Introduction}

Multiparticle production is often dubbed in terms of branching
processes. In this context, selfsimilarity and scaling are particularly
appealing concepts since they constitute an intriguingly simple and
universal organisation principle of the underlying dynamics. They are
found in a large variety of complex phenomena.

Simple discrete hierarchical branching models with selfsimilar density
fluctuations were first introduced in the description of energy
dissipation in fully developed turbulence \cite{MAN74}-\cite{MEN87}.
They
combine an independent hierarchical evolution of different branches with
a random multiplicative nature of density fluctuations. In multiparticle
processes such selfsimilar cascade models serve as paradigmatic toy
models for intermittency \cite{BIA86}-\cite{BIY90}.
Intermittency is now
experimentally well established in soft multiparticle interactions and
is now being explored in hard (perturbative) multiparton interactions.

Although these branching models describe local density fluctuations,
they are conventionally analyzed (globally) in integrated form
as is the case for the experimental data:
The underlying concepts of scaling and selfsimilar fluctuations are
revealed in a multifractal (moment) analysis \cite{FED88}, which
globally averages over the differential (local) correlation structure.
A fully differential correlation analysis itself would be a more
fundamental approach to reveal selfsimilarity and scaling and
deviations from these. However,
correlation functions of higher than second order are difficult to
measure and to visualize. Nevertheless,
we prefer to stick to a fully local
description of hierarchical branching models and their
analysis in terms of differential correlation densities.

One word about the branching models used in the course of this paper:
Alltogether they serve as (oversimplified) discrete approximations
to realistic QCD parton cascades in the perturbative regime. We do not
intend to overstress this relation, but rather prefer to study new
correlation techniques demonstrated by relatively simple branching
models. --
We restrict the discussion to one-dimensional discrete branching
processes only. Higher dimensional branching processes can be treated in
the same way as the one-dimensional case and do not change the
conclusions drawn. Discreteness which is introduced into the branching
process by allowing branching only for discrete
``time" steps of the cascade, is also not a fundamental shortcoming
compared to continuous processes as long as the step width stays small.
We are aware that a complete continuous formulation of the
branching process needs a functional approach.

In this paper we essentially convey three messages. First, we generalize
the univariate (moment) generating function to a multivariate
(correlation density) generating function and derive its evolution
equation. Message two unveils a
secret unnoticed so far in moment (multifractal) analysis, but which
is essential for comparison of model predictions with data: For
cascade models with global density fluctuations on top of local density
fluctuations, a careful distinction has to be made between moments
obtained from a (theoretical) forward evolution of the cascade and
moments obtained from an (experimental) backward analysis. Last not
least, as message three
we propose a choice of a clever basis for the representation of
correlation structures for hierarchically organized stochastic
processes: wavelets. They lead to a
tremendous simplification and directly unveil information of interest,
which is otherwise scattered in the many dimensions of phase space.

The call of wavelets \cite{MAL89}-\cite{KAI94} needs some further
introductory remarks:
The whole wavelet basis is constructed from dilations and
translations of one single ``mother" wavelet and represents
a selfsimilar and orthogonal basis. Thus we would expect that the
wavelet transform uniquely simplifies the correlation functions of
selfsimilar processes and, in particular, quasi-diagonalizes the
covariance matrix. This has been
demonstrated in ref.\ \cite{FLA92} for fractional Brownian motion and
in \cite{GRE94} for the $p$-model
\cite{MEN87}, which is the simplest tractable random cascade model.
Still more can be learned from the wavelet transform.

To estimate correlation patterns, the human brain follows a strategy
different from the standard analysis of correlations: it organizes
particles accumulated
in densely populated regions into (hard to quantify) ``clumps"
or ``clusters" and unpopulated regions into ``voids". If one looks
closer
into a particular ``clump" it may (or may not) again be organized into
``clusters" and ``voids", but now with respect to the higher (smoothed)
background density of the bigger ``parent-clump", and so on. --
Due to its multiscale localization property the
wavelet transformation dissects a (random) signal into contributions
from different scales and thus localizes small and large scale
structures separately. This feature is the reason why
wavelets are so successful in signal analysis \cite{MAL89}.

Wavelet correlations provide statistical
information about further subclustering of certain
structures like clusters, filaments or voids. Correlations between small
subclusters living inside larger clusters or voids are naturally
revealed in the wavelet transformed correlations, as are
correlations between voids and filaments. Loosely speaking, we
denote wavelet correlations as ``clump" correlations.

In this publication we study conventional and
wavelet correlations of one-dimensional
hierarchically organized branching processes.
For those readers only interested in one
or the other part, we have treated
conventional and wavelet correlations in separated sections.

In section 2 we define a class of discrete hierarchical
branching models. We distinguish between the concepts of deterministic
and random branching, which describe a zero lifetime cascade or a
cascade with finite lifetime respectively. An evolution equation for
the multivariate
generating function of the correlation densities is presented, which
allows one to calculate the correlation densities recursively. This
procedure is exemplified for the so-called $p$-model \cite{MEN87},
$\alpha$-model \cite{SCH85}, the $p$-model with random branching and
a QCD-motivated cascade model \cite{BRA94,MEU92}.
In an additional subsection we emphasize the conceptual distinction
between an evolution (= theoretical) and a
backward (= experimental) moment analysis.

The concept of wavelet correlations is introduced in section 3. For
the same models we discuss the
simplified and compressed structure of the wavelet
correlations of all orders. On the basis of the $p$-model we stress the
relationship of higher order wavelet correlations with clump
correlations. This interpretation is further illustrated
as we discuss the wavelet multiresolution analysis of
configurations belonging to two-dimensional branching processes.

In section 4, which represents the conclusions, we give an outlook
on future applications of the wavelet correlations in
multiparticle processes and other fields in physics.

\section{Hierarchical cascades \newline
         -- conventional correlation formalism}

We study a class of discrete hierarchical cascade models with what we
call random or deterministic branching and restrict ourselves to the
one-dimensional case
only. From a general evolution equation for the generating
function we calculate recursion relations for conventional correlation
densities. As specific examples, the correlation structure of the
frequently used $p$-model \cite{MEN87,LIP89} and
$\alpha$-model \cite{SCH85,BIA86,OCH90} as well as the $p$-model with
random branching and the QCD-motivated cascade model of ref.\
\cite{BRA94,MEU92}
are analysed in more detail. We exhibit the
importance of the conceptual distinction between moments referring to
the forward evolution of the cascade and moments obtained from a
backward analysis.

\subsection{Hierarchical cascading processes with random branching}

We illustrate the topological structure of a hierarchical cascading
process
with random branching in Fig.1. Starting from a single trunk, the tree
might branch into a left and a right branch at the first cascade step
with probability $\tilde{p}$; with  probability
$1-\tilde{p}$ the tree might not branch at the first cascade step. At
the second cascade step, every single branch that has formed so far
may again split into a left and a right branch with probability
$\tilde{p}$ or remains unsplit with probability $1-\tilde{p}$. This
prescription is repeated at each further cascade step. For lack of a
better terminology we associate {\em random branching} with the case
$\tilde{p} < 1$ and term the case $\tilde{p} = 1$ {\em deterministic
branching}.

In addition to the topological structure of the branching process, the
spatial structure is specified by the following rules (see Fig.~2):
The origin of the cascading tree corresponds without any loss of
generality to an interval $[0,1]$ with uniform (energy) density
$\epsilon$ normalized to one, $\epsilon_0^{(0)} = 1$.
If the
interval does not split with probability $1-\tilde{p}$ at the  first
cascade step, the energy density will remain the same as before. If, on
the other hand, the interval does branch into two halves with
probability
$\tilde{p}$, one part of the original energy $E=1$, namely
$E_L = \frac{q_L}{2}E = \frac{q_L}{2}$,
goes to the left subinterval and another part, namely
$E_R = \frac{q_R}{2}$,
goes to the right interval.
The weights $q_L$ and $q_R$ are random variables in the range
$0 \le q_L, q_R \le 2$ for both random and determinisitic branching
and follow a joint probability distribution $p(q_L,q_R)$, which
is often called a ``splitting function"
or ``splitting kernel". Depending on the splitting function
$p(q_L,q_R)$, energy may or may not be
conserved at one branching step, so that $E_L+E_R=E$ needs not be
fulfilled.
The energy densities of the resulting left and right subintervals are
$\epsilon_0^{(1)} = q_L$ and
$\epsilon_1^{(1)} = q_R$ respectively.
-- This prescription is repeated for all subsequent cascade steps
$j$. Whenever a subinterval splits into
two halves, the energy density of the left half subinterval is $q_L$
times the previous energy density,
$\epsilon_{2k}^{(j_b+1)} = q_L \epsilon_k^{(j_b)}$,
and, correspondingly, the energy density of the right half subinterval
is
$\epsilon_{2k+1}^{(j_b+1)} = q_R \epsilon_k^{(j_b)}$,
where $q_L$ and $q_R$ always follow the same splitting function.
The upper index of $\epsilon_k^{(j_b)}$ indicates the
number of branchings $j_b$ that occurred on the way to a
subinterval and
should not be confused with the number of cascade steps $j$, whereas the
lower index numbers the subinterval given at the branching scale $j_b$.
If the subinterval does not split, the energy density remains the same
as before. -- This energy curdling is shown in Fig.~2 for one possible
realization. Once the last cascade step $j=J$ is reached, the resulting
energy density distribution is resolved on this finest scale with $2^J$
subintervals (= bins). The energy densities $\epsilon_k^{(J)}$,
k=0,\ldots,$2^J-1$, of neighbouring bins then may be identical, whenever
a nonbranching has occurred during previous stages of the cascade.

In particle physics, cascade models based on continuous random variables
$\epsilon_k$ as discussed above are discretized by means of the Poisson
transform mechanism \cite{BIA86}. After the last cascade step $J$, the
energy density $\epsilon_k^{(J)}$ is replaced by a discrete particle
number $n_k$ which is drawn from a Poissonian with a mean proportional
to $\epsilon_k^{(J)} / 2^J$. The necessary transition from ordinary to
factorial correlation densities is well known, has no impact on our
results and will not be further discussed in this paper.
As has been demonstrated in ref.\ \cite{GRE94}
this also holds for the transition from ordinary to factorial
wavelet correlation densities.

\subsection{Evolution equation for conventional correlation densities}

The cascading process described in the previous subsection is
selfsimilar by construction: At each cascade step the same branching
prescription is applied as in the previous steps. As shown in later
sections this selfsimilar construction does not necessarily imply
perfect scale invariance of correlation functions. Nevertheless, a
detailed analysis of correlation densities is the only means to obtain
local information on the (selfsimilar) branching structure of a cascade
process.

The bin correlation densities are defined as the product of energy
densities contained in various bins, averaged or sampled over
all possible configurations. For example, one-, two- and three-bin
correlation densities read
\begin{eqnarray}
\label{bincorr}
  \rho_{k_1}^{(J)}
    & = &  \langle \epsilon_{k_1}^{(J)} \rangle
           \quad ,  \nonumber  \\
  \rho_{k_1k_2}^{(J)}
    & = &  \langle \epsilon_{k_1}^{(J)}
           \epsilon_{k_2}^{(J)} \rangle
           \quad ,  \\
  \rho_{k_1k_2k_3}^{(J)}
    & = &  \langle \epsilon_{k_1}^{(J)} \epsilon_{k_2}^{(J)}
           \epsilon_{k_3}^{(J)} \rangle
           \quad ;  \nonumber
\end{eqnarray}
the brackets $\langle \ldots \rangle$ indicate the averaging over all
configurations. The indices $k_i$ run from $0$ to $2^J-1$ and
represent the number of different bins at the finest resolution scale
$J$. The correlation densities are most easily determined once the
corresponding (characterisitic) generating function
\begin{equation}
\label{genfunc}
  Z^{(J)}\left[\vec{\lambda}^{(J)}\right]
    =  \left\langle \exp \left( i \sum_{k=0}^{2^J-1} \lambda_k^{(J)}
                                \epsilon_k^{(J)}
                         \right) \right\rangle
\end{equation}
is known; they follow by taking appropriate derivatives with
respect to the $\lambda_k^{(J)}$:
\begin{eqnarray}
\label{derivgen}
  \rho_{k_1}^{(J)}
    & = &  \frac{1}{i} \left.
           \frac{\partial Z^{(J)}\left[\vec{\lambda}^{(J)}\right]}
                {\partial\lambda_{k_1}^{(J)}}
           \right|_{\vec{\lambda}^{(J)}=0}
           \quad ,  \nonumber  \\
  \rho_{k_1k_2}^{(J)}
    & = &  \frac{1}{i^2} \left.
           \frac{\partial^2 Z^{(J)}\left[\vec{\lambda}^{(J)}\right]}
                {\partial\lambda_{k_1}^{(J)}
                 \partial\lambda_{k_2}^{(J)}}
           \right|_{\vec{\lambda}^{(J)}=0}
           \quad ,  \\
  \rho_{k_1k_2k_3}^{(J)}
    & = &  \frac{1}{i^3} \left.
           \frac{\partial^3 Z^{(J)}\left[\vec{\lambda}^{(J)}\right]}
                {\partial\lambda_{k_1}^{(J)}
                 \partial\lambda_{k_2}^{(J)}
                 \partial\lambda_{k_3}^{(J)}}
           \right|_{\vec{\lambda}^{(J)}=0}
           \quad ,  \quad \ldots \; .  \nonumber
\end{eqnarray}
In the following we develop a scheme how to determine the
generating function at a given evolution scale from the previous
evolution scales. This scale-dependent recursion relation then leads
to scale-dependent recursion relations for the bin correlation
densities.

Suppose we know the generating function
$Z^{(j)}\left[\vec{\lambda}^{(j)}\right]$
after $j$ cascade steps. Then we can determine
$Z^{(j+1)}$ after $j+1$ cascade steps by a backward evolution: With
probability $(1-\tilde{p})$ there has been no branching in this backward
evolution step and with probability $\tilde{p}$ a branching into a left
and right branch has occurred, where the split energy densities have
been weighted with $q_L$ and $q_R$ respectively. This translates into
\begin{equation}
\label{genevolu}
  Z^{(j+1)}\!\left[\vec{\lambda}^{(j+1)}\right]
    =  (1-\tilde{p}) \; Z^{(j)}\!\left[\vec{\lambda}_M^{(j)}\right]
       + \tilde{p} \int\! {\rm d}q_L {\rm d}q_R \; p(q_L,q_R) \;
         Z^{(j)}\!\left[q_L\vec{\lambda}_L^{(j)}\right]
         Z^{(j)}\!\left[q_R\vec{\lambda}_R^{(j)}\right]
       \quad ,
\end{equation}
where the rules
\begin{eqnarray}
\label{lamevo1}
  \vec{\lambda}^{(j+1)}
    & = &  \left( \lambda_0^{(j+1)} , \, \lambda_1^{(j+1)}
                  , \, \ldots , \, \lambda_{2^{j+1}-1}^{(j+1)}
           \right) \quad ,  \nonumber  \\
  \vec{\lambda}_L^{(j)}
    & = &  \left( \lambda_0^{(j+1)} , \, \lambda_1^{(j+1)}
                  , \, \ldots , \, \lambda_{2^j-1}^{(j+1)}
           \right) \quad ,  \nonumber  \\
  \vec{\lambda}_R^{(j)}
    & = &  \left( \lambda_{2^j}^{(j+1)} , \, \lambda_{2^j+1}^{(j+1)}
                  , \, \ldots , \, \lambda_{2^{j+1}-1}^{(j+1)}
           \right) \quad ,  \nonumber  \\
  \vec{\lambda}_M^{(j)}
    & = &  \left( \lambda_0^{(j+1)} + \lambda_1^{(j+1)}, \,
                  \lambda_2^{(j+1)} + \lambda_3^{(j+1)}, \,
                  \ldots , \,
                  \lambda_{2^{j+1}-2}^{(j+1)} +
                    \lambda_{2^{j+1}-1}^{(j+1)}
            \right)
\end{eqnarray}
specify the spatial splitting structure of the branching process.
The splitting function $p(q_L,q_R)$ reflects the probability that
random
weights $q_L$ and $q_R$ are assigned to the energy densities of the left
and right branch respectively and is normalized to one, i.e.\
$\int p(q_L,q_R) {\rm d}q_L {\rm d}q_R = 1$.
The nonlinear backward evolution equation (\ref{genevolu}) allows one to
express
$Z^{(j+1)}$ in terms of $Z^{(j)}$, which in turn can be expressed
in terms of $Z^{(j-1)}$, and so on until the roughest scale $j=0$ is
reached, where
\begin{equation}
\label{evostart}
  Z^{(0)}\!\left[\lambda_0^{(0)}\right]
    =  e^{i \lambda_0^{(0)}}
       \quad .
\end{equation}
The evolution equation (\ref{genevolu}) has been derived in detail in
\cite{GRE94} for the case of the $p$-model with deterministic
branching.

It is instructive to rewrite Eq.(\ref{genevolu}) as
\begin{equation}
\label{contin2}
  \frac{\Delta Z^{(j)}\!\left[\vec{\lambda}\right]}{\Delta j}
    =  \frac{1}{\tau}
       \int\! {\rm d}q_L {\rm d}q_R \; p(q_L,q_R)
       \left\{
         Z^{(j)}\!\left[q_L\vec{\lambda}_L\right]
         Z^{(j)}\!\left[q_R\vec{\lambda}_R\right]
         - Z^{(j)}\!\left[\vec{\lambda}_M\right]
       \right\}  \quad ,
\end{equation}
where
$\Delta Z^{(j)}\!\left[\vec{\lambda}\right] =
 Z^{(j+1)}\!\left[\vec{\lambda}^{(j+1)}\right]
 - Z^{(j)}\!\left[\vec{\lambda}_M^{(j)}\right]$,
$\Delta j = 1$ and
$1/\tau = \tilde{p} / \Delta j$
is a branching rate. This form is a discrete analogue to the evolution
equations widely used to model QCD parton showers in the
perturbative regime \cite{CVI80}. By this analogy we are
motivated to study the mathematical structure of the
recursive discrete evolution
equation (\ref{genevolu}) in detail.

{}From Eq.~(\ref{genevolu})
we derive the recursion relations for the bin correlation densities by
taking the relevant derivatives (\ref{derivgen}) with respect to the
$\lambda_k^{(j+1)}$ taking into account the transformations
(\ref{lamevo1}). For the one-bin correlation density we deduce:
\begin{equation}
\label{onecorec}
  \rho_{k_1}^{(j+1)}
     =  (1 - \tilde{p}) \rho_{m_1}^{(j)}
        + \tilde{p} \, \overline{q} \, \rho_{m_2}^{(j)}
        \quad ;
\end{equation}
the following four cases have to be distinguished:
\begin{itemize}
  \item  $k_1 \in \{0 , \ldots , 2^j-1\} = \{ L \}$ and
         $k_1=2m$ an even number: \\
         $m_1=m=k_1/2$, $m_2=k_1$,
         $\overline{q} = \overline{q_L}
          = \int p(q_L,q_R) \, q_L \, {\rm d}q_L {\rm d}q_R$,
  \item  $k_1 \in \{ L \}$ and
         $k_1=2m+1$ an odd number: \\
         $m_1=m=(k_1-1)/2$, $m_2=k_1$,
         $\overline{q} = \overline{q_L}$,
  \item  $k_1 \in \{2^j , \ldots , 2^{j+1}-1\} = \{ R \}$ and
         $k_1=2m$ an even number: \\
         $m_1=m=k_1/2$, $m_2=k_1-2^j$,
         $\overline{q} = \overline{q_R} =
          \int p(q_L,q_R) \, q_R \, {\rm d}q_L {\rm d}q_R$,
  \item  $k_1 \in \{ R \}$ and
         $k_1=2m+1$ an odd number: \\
         $m_1=m=(k_1-1)/2$, $m_2=k_1-2^j$,
         $\overline{q} = \overline{q_R}$.
\end{itemize}
Because the splitting function $p(q_L,q_R)$ needs not be symmetric in
$q_L$ and $q_R$ we distinguish $\overline{q_L}$
and $\overline{q_R}$. For
completeness we state that $\rho_0^{(0)}=1$, which follows directly from
Eqs.\ (\ref{derivgen}) and (\ref{evostart}). -- For the two-bin
correlation density we have to distinguish two principal cases. If $k_1$
and $k_2$ belong to the same branch, i.e.\ $k_1,k_2 \in \{ L \}$ or
$k_1,k_2 \in \{ R \}$, we get
\begin{equation}
\label{twocore1}
  \rho_{k_1k_2}^{(j+1)}
     =  (1 - \tilde{p}) \rho_{m_1m_2}^{(j)}
        + \tilde{p} \, \overline{q^2} \, \rho_{m_3m_4}^{(j)}
\end{equation}
with $\overline{q^2} = \overline{q_{L}^2}
      = \int p(q_L,q_R) \, q_L^2 \, {\rm d}q_L {\rm d}q_R$
or $\overline{q^2} = \overline{q_{R}^2}$ analogously.
If $k_1$ and $k_2$ belong to different branches, i.e.\
$k_1 \in \{ L \}$, $k_2 \in \{ R \}$ or vice versa, we get
\begin{equation}
\label{twocore2}
  \rho_{k_1k_2}^{(j+1)}
     =  (1 - \tilde{p}) \rho_{m_1m_2}^{(j)}
        + \tilde{p} \, \overline{q_Lq_R} \,
          \rho_{m_3}^{(j)} \rho_{m_4}^{(j)}
\end{equation}
with $\overline{q_Lq_R}
            = \int p(q_L,q_R) \, q_L q_R \, {\rm d}q_L {\rm d}q_R$.
Several cases have to be distinguished for the indices
$k_1,k_2 \rightarrow m_1,m_2,m_3,m_4$; they follow from the
transformations (\ref{lamevo1}) and will not be
further specified. Again, for completeness,
$\rho_{00}^{(0)} = 1$. -- Although the strategy how to
calculate bin correlation densities of arbitrary order should be clear,
we still want to mention three-bin correlations. Two principal
cases have to be taken into account: if $k_1,k_2,k_3$ all belong to the
same branch we have
\begin{equation}
\label{thrcore1}
  \rho_{k_1k_2k_3}^{(j+1)}
     =  (1 - \tilde{p}) \rho_{m_1m_2m_3}^{(j)}
        + \tilde{p} \, \overline{q^3} \, \rho_{m_4m_5m_6}^{(j)}
\end{equation}
with $\overline{q^3} = \overline{q_L^3}
            = \int p(q_L,q_R) \, q_L^3 \, {\rm d}q_L {\rm d}q_R$
or $\overline{q^3} = \overline{q_R^3}$, respectively, and if
$k_1,k_2$ belong for example to the left branch and $k_3$ to the right
branch we have
\begin{equation}
\label{thrcore2}
  \rho_{k_1k_2k_3}^{(j+1)}
     =  (1 - \tilde{p}) \rho_{m_1m_2m_3}^{(j)}
        + \tilde{p} \, \overline{q_L^2q_R} \,
          \rho_{m_4m_5}^{(j)} \rho_{m_6}^{(j)}
        \quad .
\end{equation}
The first value
for the iteration is given by $\rho_{000}^{(0)} = 1$.

\subsection{Specific cascade models}

The recursion relations (\ref{onecorec})-(\ref{thrcore2}) for the bin
correlation densities are quite general. Models of hierarchical
random branching
processes differ in the branching probability and the splitting
function, the latter giving rise to different splitting moments
$\overline{q_L^mq_R^n}$.
In this subsection we concentrate first on two specific cascade models
with determinstic branching, the $p$-model
\cite{MEN87} and the $\alpha$-model \cite{SCH85}. They differ in
that the former conserves energy at each branching whereas the
latter does
not. Both models have been suggested to describe the intermittent energy
dissipation of fully developed turbulence; both also serve as simple
discrete approximations to multiparticle processes in high energy
$e^+e^-$ and hadron-hadron collisions \cite{BIA86}-\cite{OCH90}.
Then we discuss the effects of random branching for the case of the
$p$-model. Finally, a model in the context of perturbative QCD parton
cascades \cite{BRA94,MEU92} is analyzed.

\subsubsection{$p$-model}

For the $p$-model \cite{MEN87} the branching probability is equal to
$\tilde{p}=1$; we call this deterministic branching. With every
branching the energy splits into two unequal portions governed by the
splitting factors $q_L/2=(1+\alpha)/2$ and $q_R/2=(1-\alpha)/2$
or vice versa; both
possibilities occur with probability $1/2$. The splitting function of
the $p$-model then reads:
\begin{equation}
\label{splitp}
  p(q_L,q_R)
    =  \frac{1}{2} \Big(
       \delta(q_L-(1+\alpha)) + \delta(q_L-(1-\alpha))
       \Big) \delta(q_L+q_R-2)
       \quad .
\end{equation}
The last $\delta$-function in this expression guarantees that energy is
conserved in the splitting process. Note that $q_L$ and $q_R$ are
weight factors for the energy densities and not the energies, so that
$q_L+q_R=2$ and not equal to one. The moments of the splitting function
which enter in the recursion relations for the correlation densities
(\ref{onecorec})--(\ref{thrcore2}) now become:
\begin{eqnarray}
\label{splitmom}
  \overline{q_L}
    & = &  \overline{q_R} \;
           =  \int_0^2 \! p(q_L,q_R) \, q_L \, {\rm d}q_L {\rm d}q_R \;
           =  1
           \quad ,  \nonumber  \\
  \overline{q_L^2}
    & = &  \overline{q_R^2} \;
           =  \; 1+\alpha^2
           \quad , \quad\quad \;\,
           \overline{q_Lq_R} \; = \; 1-\alpha^2
           \quad ,  \nonumber  \\
  \overline{q_L^3}
    & = &  \overline{q_R^3} \;
           =  \; 1+3\alpha^2
           \quad , \quad\quad
           \overline{q_L^2q_R} \; = \;
           \overline{q_Lq_R^2} \; = 1-\alpha^2
           \quad ;
\end{eqnarray}
they determine the scaling indices of the correlation densities.

{}From the recursion
relation (\ref{onecorec}) it follows that the bin correlation densities
of first order are equal to
\begin{equation}
\label{rhobin1}
  \rho_{k_1}^{(J)}  =  1   \quad ,
\end{equation}
because the first order splitting moments in Eq.~(\ref{splitmom})
are equal to one.
This is just another way of saying that on average the energy density
at scale $J$ is equal to the energy density at scale $J=0$. For
second order bin correlation densities we find from Eqs.\
(\ref{twocore1}), (\ref{twocore2}) and (\ref{splitmom})
\begin{equation}
\label{rhobin2}
  \rho_{k_1k_2}^{(j+1)}
    =  \left\{  \begin{array}{ll}
       (1+\alpha^2) \rho_{k_1k_2}^{(j)}
       &  , {\rm if} \; k_1,k_2\in\{ L\}  \quad , \\
       (1+\alpha^2) \rho_{k_1-2^j,k_2-2^j}^{(j)}
       &  , {\rm if} \; k_1,k_2\in\{ R\}  \quad , \\
       (1-\alpha^2)
       &  , {\rm if} \; k_1\in\{ L\} \; {\rm and} \; k_2\in\{ R\}
          {\rm \; or \, vice \, versa.}
       \end{array}  \right.
\end{equation}
The power-law-like singularity towards the diagonal can be
seen clearly in Fig.~3a.
The closer the bins are together topologically the more they share a
common history
and the stronger they are correlated. This is a characteristic feature
of selfsimilarity. We shed some more light onto Eq.\ (\ref{rhobin2}),
once we compare for example
$\rho_{00}^{(J)}=\rho_{11}^{(J)}=(1+\alpha^2)^{J}$ and
$\rho_{01}^{(J)}=\rho_{10}^{(J)}=(1+\alpha^2)^{J-1}(1-\alpha^2)$;
they
differ by a factor $(1+\alpha^2)/(1-\alpha^2)$.
This represents an anticorrelation
caused by energy conservation: $\epsilon_0^{(j+1)}$ and
$\epsilon_1^{(j+1)}$ of one possible configuration differ insofar that
only at the very last cascade step different energy-conserving weights
$(1+\alpha)$ and $(1-\alpha)$ were assigned to them.

\subsubsection{$\alpha$-model}

Like the $p$-model, the $\alpha$-model is a deterministic cascade, so
that $\tilde{p}=1$. The
splitting function factorizes and is now given as
\begin{eqnarray}
\label{splitalp}
  p(q_L,q_R)
    & = &  \Big( p_1 \delta(q_L-(1-\alpha)) + p_2 \delta(q_L-(1+\beta))
           \Big)   \nonumber  \\
    &   &  \cdot
           \Big( p_1 \delta(q_R-(1-\alpha)) + p_2 \delta(q_R-(1+\beta))
           \Big)
           \quad ,
\end{eqnarray}
where the splitting parameters $\alpha$, $\beta$ are positive real
numbers. The probabilities
$p_1=\beta/(\alpha+\beta)$ and $p_2=\alpha/(\alpha+\beta)$ are
chosen such that $p_1+p_2=1$ and $p_1(1-\alpha)+p_2(1+\beta)=1$. In
contrast to the $p$-model, energy needs not be conserved in every
cascade step; for example the
energy density may split into a left part with weight factor
$(1-\alpha)$
and into a right part with the same weight factor $(1-\alpha)$
with a probability $p_1^2$.
Energy is however conserved on average since $\overline{q_{L/R}}=1$.
For the
splitting moments appearing in eqs.\ (\ref{onecorec})-(\ref{thrcore2})
we derive from (\ref{splitalp})
\begin{eqnarray}
\label{splitmoma}
  \overline{q_L}
    & = &  \overline{q_R}
           =  1   \quad ,  \nonumber  \\
  \overline{q_L^2}
    & = &  \overline{q_R^2}
           = 1 + \alpha\beta  \quad ,  \quad\quad\quad\quad\quad\quad
                                       \quad
           \overline{q_Lq_R}
           =  1   \quad ,  \nonumber  \\
  \overline{q_L^3}
    & = &  \overline{q_R^3}
           = 1 + \alpha\beta(3-\alpha+\beta)  \quad ,  \quad\quad
           \overline{q_L^2q_R}
           = \overline{q_Lq_R^2}
           = 1 + \alpha\beta   \quad .
\end{eqnarray}

With the first order splitting moments $\overline{q_L}$,
$\overline{q_R}$ and the branching probability $\tilde{p}=1$
we derive from the recursion relation (\ref{onecorec})
\begin{equation}
\label{rhobin1a}
  \rho_{k_1}^{(J)}  =  1   \quad .
\end{equation}
On average the energy density contained in one bin is equal to the
original energy density.

According to the
recursion relations (\ref{twocore1}), (\ref{twocore2}) and the
corresponding splitting moments (\ref{splitmoma}) we derive for the
second order bin correlations
\begin{equation}
\label{rhobin2a}
  \rho_{k_1k_2}^{(j+1)}
    =  \left\{  \begin{array}{ll}
       (1+\alpha\beta) \rho_{k_1k_2}^{(j)}
       &  , {\rm if}\; k_1,k_2\in \{ L\}\; {\rm or}\, \ldots  \\
       \quad 1
       &  , {\rm if}\; k_1\in \{ L\}\; {\rm and}\; k_2\in \{ R\}\;
          {\rm or \; vice \; versa},
       \end{array}  \right.
\end{equation}
which are visualized in Fig.~3b.
Apparently there is no big difference to the corresponding
$p$-model relations (\ref{rhobin2}): A powerlaw scaling towards the
diagonal emerges, which is a clear evidence of the underlying
selfsimilarity in the $\alpha$-model cascade. The
difference to the $p$-model correlations lies in the absence of
anticorrelations due to the violation of energy conservation.
As soon as two different bins start to belong to different
branches during the evolution of the cascade they become completely
uncorrelated. We will see later on, that this missing anticorrelation is
responsible for the deviation from perfect scaling behaviour of the
$\alpha$-model in a moment (multifractal) analysis.

\subsubsection{$p$-model with random branching}

In the previous subsections we have discussed the $p$- and
$\alpha$-model
as two representatives of deterministic branching processes.
Although these hierarchical models are intended as simple
discrete approximations to multiparticle cascading
\cite{BIA86}-\cite{OCH90}, they might not be that realistic.
Multiparticle branchings have to be somehow ``undeterministic": With a
certain probability a particle might not decay into two other particles
at every step of the cascade; so to say the particle has a certain
lifetime. Therefore we incorporate a nonbranching part into the
hierarchical cascade process as described in
sections 2.1 and 2.2. In fact, the branching rate $1/\tau$ introduced
in the evolution equation (\ref{contin2})
can be interpreted as the inverse of the particle's lifetime once the
scale evolution parameter $j$ is identified with time.
With this motivation we study the effect of random
branching on selfsimilar cascading.

As in the deterministic $p$-model, we choose the splitting function
(\ref{splitp}) with the corresponding splitting moments
(\ref{splitmom}). The branching probability $\tilde{p}$ is now chosen
in between $0 < \tilde{p} < 1$; the case $\tilde{p} = 1$ defines
deterministic branching and the case $\tilde{p} =0$ describes no
branching at all. The recursion relation (\ref{onecorec}) for the
one-bin correlation density has as solution again
$\rho_{k_1}^{(J)} = 1$. The solution of the recursion
relations (\ref{twocore1}) and (\ref{twocore2}) for the two-bin
correlation density is depicted in Fig.\ 3c; there is an approximate
power-law rise far from the diagonal, but the closer the diagonal is
approached the more the rise turns into a plateau. The explanation for
this
behaviour lies in the following: With an increasing number of cascade
steps the probability that between neighbouring bins no
branching has occurred becomes overwhelming.
Then the energy densities of neighbouring bins
are the same, so that the two-bin correlation density near the diagonal
becomes constant.

\subsubsection{QCD-motivated cascade model}

As a generalization of the $p$- and $\alpha$-model, so-called pseudo
QCD-cascade models have been discussed in the context of
multiparticle dynamics \cite{CHU90}. Recently Brax, Meunier and
Peschanski \cite{BRA94} have suggested a discrete hierarchical branching
model, which reproduces the same multiplicity moments
as more elaborate gluon jet cascade models within perturbative QCD
\cite{BRA94},\cite{OCH92}-\cite{DOK93}.
In this scenario the interval splitting at each branching step is
interpreted as a partitioning in opening angle. The simplicity of the
QCD motivated model of refs.\ \cite{BRA94,MEU92} permits a
full analytical treatment in terms of (local) correlation densities
besides (global) moments.

The branching probability is set to $\tilde{p}=1$, reflecting the "zero
lifetime" of the virtual gluons. The splitting function reads
\begin{equation}
\label{qcdsplit}
  p(q_L,q_R) = \frac{1}{2} \left[
               \delta(q_L-2) \left( \delta(q_R) + \left(
                 \frac{\gamma}{q_R} \right)_{\! +} \right)
               +
               \left( \delta(q_L) + \left( \frac{\gamma}{q_L}
                 \right)_{\! +} \right) \delta(q_R-2)
               \right]
\end{equation}
and describes the decay of a hard parent gluon into one hard and one
soft daughter gluon. The quantity
$(1/q)_+ = \lim_{\beta \rightarrow 0} \left( (1/q)
 \Theta(q-2\beta) + \delta(q-2\beta) \ln \beta \right)$
indicates a regularization prescription \cite{FIE89}. The parameter
$\gamma = \gamma_0 \ln 2 = \sqrt{6\alpha_s/\pi} \ln 2$
is proportional to the fixed strong coupling constant
$\sqrt{\alpha_s}$. With (\ref{qcdsplit}) the splitting moments become
\begin{eqnarray}
\label{qcdspmom}
  \overline{q_L}
    & = &  \overline{q_R} \;
           =  \int_0^2 \! {\rm d}q_L {\rm d}q_R \;
              p(q_L,q_R) \, q_L
              \;
           =  1 + \gamma
           \quad ,  \nonumber  \\
  \overline{q_L^2}
    & = &  \overline{q_R^2} \;
           =  \; 2 (1+\frac{\gamma}{2})
           \quad , \quad\quad\quad
           \overline{q_Lq_R} \; = \; 2 \gamma
           \quad ,  \nonumber  \\
  \overline{q_L^3}
    & = &  \overline{q_R^3} \;
           =  \; 4 (1+\frac{\gamma}{3})
           \quad , \quad\quad\quad
           \overline{q_L^2q_R} \; = \;
           \overline{q_Lq_R^2} \; = 6 \gamma
           \quad .
\end{eqnarray}
This sets the stage for the conventional correlation densities
(\ref{onecorec})-(\ref{thrcore2}).

For the first order correlation density we find
\begin{equation}
\label{qcdcor1}
  \rho_{k_1}^{(J)}
    =  (1 + \gamma)^J
       \quad .
\end{equation}
The splitting function (\ref{qcdsplit}) does not conserve energy; it
produces additional energy in every cascade step. Here, we should
think of ``energy" as ``multiplicity", so that (\ref{qcdcor1}) describes
the average production of multiplicity during the evolution of the
cascade.

With the help of the recursion relations (\ref{twocore1}) and
(\ref{twocore2}), the relevant splitting moments (\ref{qcdspmom}) and
the result (\ref{qcdcor1}) we calculate for the bin correlation
densities of second order
\begin{equation}
\label{qcdcor2}
  \rho_{k_1k_2}^{(j+1)}
    =  \left\{  \begin{array}{ll}
       (2+\gamma) \rho_{k_1k_2}^{(j)}
       &  , {\rm if}\; k_1,k_2\in \{ L\}\; {\rm or}\, \ldots  \\
       2 \gamma (1+\gamma)^{2j}
       &  , {\rm if}\; k_1\in \{ L\}\; {\rm and}\; k_2\in \{ R\}\;
          {\rm or \; vice \; versa}.
       \end{array}  \right.
\end{equation}
The reduced second order correlation density
$\rho_{k_1k_2} / \rho_{k_1} \rho_{k_2}$
is shown in Fig.\ 3d. Note that for the $p$- and
$\alpha$-model with and without random branching it was superfluous to
distinguish between a reduced and non-reduced correlation density
because there $\rho_k=1$. This is not anymore the case for the
QCD-motivated cascade model treated here.

{}From Eq.\ (\ref{qcdcor2}) it follows with $k_1=0$ that
$\rho_{0,k}/\rho_{0,2k} = (2+\gamma)/(1+\gamma)^2$
for $k \neq 0$ and
$\rho_{0,0}/\rho_{0,1} = (2+\gamma)/2\gamma$. This implies a scaling
$[ (2+\gamma) / (1+\gamma)^2 ]^j$
towards the diagonal except for the last step. It is the same scaling
we encounter in a (forward) normalized moment analysis in section
2.4.2. Due to the pronounced ``non-scaling" peak along the diagonal
and the non-matching anticorrelations, the correct
(backward) moments will deviate from this perfect scaling.

\subsection{Moments: forward evolution vs. backward analysis}

Contrary to common belief, the $\alpha$-model does not exhibit a perfect
power law in a (backward) moment analysis in spite of its strictly
selfsimilar (forward) evolution. This statement also holds for the
QCD-motivated cascade model discussed above.
In order to unveil this hidden secret
we recall the standard procedure of a moment analysis as proposed in
ref.\ \cite{BIA86}.

For example, the second order moment $M_2(J,J)$ is defined by
\begin{equation}
\label{mom1}
  M_2(J,J)
    =  \frac{1}{2^J} \sum_{k=0}^{2^J-1}
       \left\langle ( \epsilon_k^{(J)} )^2 \right\rangle
    =  \frac{1}{2^J} \sum_{k=0}^{2^J-1}
       \rho_{kk}^{(J)}
    =  \rho_{00}^{(J)}
       \quad ,
\end{equation}
where the first index $j_1$ in the nonstandard, double-indexed notation
$M_2(j_1,j_2)$ denotes the total number of cascade steps $J$ after a
forward evolution whereas the second index $j_2 \leq j_1$ represents the
resolution scale ($\sim$ log(bin size)), at which the cascade is
analyzed in backward direction. Note that
the bin correlation densities $\rho_{kk}^{(J)}$ of the $p$- and
$\alpha$-model as well as of the QCD-motivated cascade model
are independent of the
shift index $k$. Now we average over two neighbouring bins and arrive at
the moment
\begin{equation}
\label{mom2}
  M_2(J,J-1)
    =  \frac{1}{2^{J-1}} \sum_{k=0}^{2^{J-1}-1}
       \left\langle \left( \frac{1}{2}
       ( \epsilon_{2k}^{(J)} + \epsilon_{2k+1}^{(J)} ) \right)^2
       \right\rangle
    =  \frac{1}{2} ( \rho_{00}^{(J)} + \rho_{01}^{(J)} )
    =  \frac{1}{2} ( M_2(J,J) + \rho_{01}^{(J)} )
       \quad ,
\end{equation}
resolved at the scale $j_2=J-1$.
Then, we continue to average over four, eight, \ldots neighbouring bins
and arrive at the following recursion relation for the moments:
\begin{equation}
\label{mom3}
  M_2(J,J-j-1)
    =  \frac{1}{2} \left( M_2(J,J-j) + \rho_{0,2^{j}}^{(J)} \right)
       \quad .
\end{equation}

\subsubsection{$p$- and $\alpha$-model}

For the $p$-model we realize with the help of (\ref{rhobin2}) that
\begin{equation}
\label{momp}
  p{\rm -model: } \quad\quad\quad
  M_2(J,j)
    =  M_2(j,j)
    \equiv
       \rho_{kk}^{(j)}
    =  (1+\alpha^2)^{j}   \quad .
\end{equation}
The moment obtained after $J$ cascade steps, averaged over bins of size
$2^{J-j}$ and then again averaged over $2^j$ blocks of bins
is equal to the diagonal element of the bin correlation
density obtained after $j$ cascade steps.

This does not hold anymore
for the $\alpha$-model: From (\ref{mom3}) and (\ref{rhobin2a}) we derive
\begin{equation}
\label{moma}
  \alpha{\rm -model: } \quad\quad\quad
  M_2(J,j)
    =  \frac{1}{(1-\alpha\beta)} (1+\alpha\beta)^{j}
        \left\{ 1 - \alpha\beta \left( \frac{1+\alpha\beta}{2}
        \right)^{J-j} \right\}  \quad .
\end{equation}
In Fig.~4 we summarize the various scaling aspects of the moments
$M_2(J,j)$.
Obviously, the backward analyzed moment $M_2(J,j)$ is not equal to the
bin correlation density
$\rho_{kk}^{(j)} = (1+\alpha\beta)^{j} = M_2(j,j)$ obtained
after $j$ forward $\alpha$-model cascade steps. The reason for this lies
in the fact that energy is not conserved in the $\alpha$-model; an
energy density $\epsilon_k^{(j)}$ obtained after $j$ cascade steps might
split into two energy densities
$\epsilon_{2k}^{(j+1)}=(1+\beta)\epsilon_{k}^{(j)}$ and
$\epsilon_{2k+1}^{(j+1)}=(1+\beta)\epsilon_{k}^{(j)}$ in the next
cascade step. But then
$\epsilon_k^{(j)} \neq \frac{1}{2} ( \epsilon_{2k}^{(j+1)} +
 \epsilon_{2k+1}^{(j+1)} )$
so that
$M_2(j+1,j) \neq \rho_{kk}^{(j)} = M_2(j,j)$.

Here it is really important to distinguish between
the moment analysis in evolution direction of the cascade,
i.e.\ $m_2(j)=M_2(j,j)$ in dependence of $j$, and in backward
direction after $J$ cascade steps, i.e.\ $M_2(J,j)$ in dependence
of $j$. An experimental moment analysis is always based on moments
like $M_2(J,j)$: every configuration (event) is analyzed
backwards, given a fully developed cascade at final scale $J$.
{}From these considerations about the $\alpha$-model we learn,
that the (observed) moments $M_2(J,j)$ may not be naively identified
with the forward evolution moments $m_2(j)$. This is in
contrast with the results of the $p$-model cascade, where the
anticorrelations due to energy conservation make a distinction
between an evolution and a backward analysis unnecessary.

\subsubsection{QCD-motivated cascade model}

{}From the (backward) moment recursion relation (\ref{mom3}) and
Eq.~(\ref{qcdcor2}) we derive for the QCD-motivated cascade model
\begin{equation}
\label{qcdm2}
  M_2(J,j)
    =  2 \frac{(1+\gamma)^{2J}}{3+2\gamma}
       \left( \frac{2+\gamma}{(1+\gamma)^2}
       \right)^j
       +
       \frac{1+2\gamma}{3+2\gamma}
       \left( 1 + \frac{\gamma}{2} \right)^J 2^j
       \quad .
\end{equation}
This is an unnormalized moment. For a proper normalization we
have to divide by the square of the density
$(\rho_0^{(J)})^2 = (1+\gamma)^{2J}$
from Eq.~(\ref{qcdcor1}); this normalization does not affect the scale
$j$ dependence.

We observe that the above expression consists  of two
terms with different scaling behaviour. The first term scales with a
factor we have already noticed in the second order correlation densities
(\ref{qcdcor2}), whereas the second term scales trivially in
the factor 2. This can be explained in the following way: The
second order correlation density (\ref{qcdcor2}) can be decomposed into
a correlation matrix with exact anticorrelations and an additional
purely diagonal correlation matrix; the former gives rise to the first
term in (\ref{qcdm2}) with anomalous scaling, whereas the latter
accounts for the trivial scaling in powers of 2 of the second term.

Note in this context, that the expression for the second order moment
(\ref{moma}) of the $\alpha$-model can also be cast in this form: one
term scales with the factor $(1+\alpha\beta)$, whereas the second term
scales with the factor 2. In contrast to the $\alpha$-model result, the
second term in (\ref{qcdm2}) for the QCD-motivated cascade model comes
with a positive sign. As a consequence, in a log-log plot the curve for
$M_2$ bends upwards for the fine resolution scales, i.e.\ large $j$
values, and deviates from the perfect scaling shown at the rougher
resolution scales, i.e.\ small $j$ values, where the first term in
(\ref{qcdm2}) dominates over the second one. This result is illustrated
in Fig.~5; compare also again with the corresponding $\alpha$-model
result shown in Fig.~4.

We emphasize again that the (backward) moment $M_2(J,j)$ of
(\ref{qcdm2}) is the relevant moment to be compared to an experimental
analysis. In theory, the normalized moments $m_q(j)$ in
forward evolution of the cascade are calculated in most cases and
compared directly to experimental data. For example, the
corresponding normalized moment of second order is defined as
\begin{equation}
\label{qcdfm2a}
  m_2(j)
    =  \frac{\rho_{kk}^ {(j)}}{(\rho_k^ {(j)})^2}
       \quad .
\end{equation}
Here the diagonal elements of the correlation densities enter for a
cascade with $j$ cascade steps; the cascade is {\em not} evolved down to
the finest scale $J$ and then resolved backwards on a rougher resolution
scale $j$. With the results (\ref{qcdcor1}) and (\ref{qcdcor2}) we find
for the QCD-motivated cascade model the scaling law
\begin{eqnarray}
\label{qcdfm2b}
  m_2(j)
    & = &  \left( \frac{2+\gamma}{(1+\gamma)^2}
           \right)^j
           \nonumber \\
    & = &  \left( 2^{-j} \right)^{-\log_2 (2+\gamma) +
                                  2\log_2 (1+\gamma)}
           \nonumber \\
    & \approx &
           \left( 2^{-j} \right)^{-(1-\frac{3}{2}\gamma_0)}
           \quad\quad\quad ({\rm for} \; \gamma_0 \ll 1).
\end{eqnarray}
It shows the same anomalous scaling as the first term in the
corresponding (backward) moment (\ref{qcdm2}) and is also illustrated in
Fig.~5. The exponent $\frac{3}{2}\gamma_0$ appearing in the last
expression is identified with the Renyi dimension $D_2$.

For the
(forward) normalized moments of higher order the calculation is
straightforward; we state the outcome without giving further
details:
\begin{eqnarray}
\label{qcdfmqa}
  m_q(j)
    & = &  \left( 2^{-j}
           \right)^{-\log_2 \left[ 2^{q-1}(1+\frac{\gamma}{q})
                            \right]
                    + q\log_2 (1+\gamma)}
           \nonumber \\
    & \approx &
           \left( 2^{-j} \right)^{-(q-1)(1-D_q)}
           \quad\quad\quad ({\rm for} \; \gamma_0 \ll 1)
\end{eqnarray}
with
\begin{equation}
\label{qcdfmqb}
  D_q  =  \frac{q+1}{q} \gamma_0
          \quad .
\end{equation}
These scaling exponents for the (forward) evolution moments have been
derived previously in the context of a perturbative QCD gluon cascade
\cite{BRA94},\cite{OCH92}-\cite{DOK93}. Note again that it is not these
unobservable
(forward) evolution moments which should be
compared to experimental data, but only experimentally accessible
(backward) moments.

\section{Wavelet correlations in hierarchical branching processes}

The conventional correlation densities (\ref{bincorr}) characterize the
correlations between bin contents resolved at some finest scale $J$
(usually chosen somewhere near the resolution limit of the measurement
device). They do not provide a natural description in terms of larger
structures (clusters and voids). However, sometimes a choice of a clever
basis for the representation of the bin contents can directly unveil
this information of interest. The wavelet transform
\cite{MAL89}-\cite{KAI94}
appears to be such a clever change of basis functions for
hierarchically organized stochastic processes \cite{FLA92,GRE94}. The
strategy applied in this section is the following: We apply a
multiresolution analysis arising from the wavelet transform to each
realization of the branching process and study the correlations between
the wavelet amplitudes. This procedure is exemplified for the same
cascade models discussed in section 2.

\subsection{Multiresolution analysis}

In this subsection
we will briefly recite only the most important ingredients of a
wavelet-based multiresolution analysis. We will
concentrate on the Haar wavelet only, which is the simplest of all
wavelets; generalizations to other wavelets are straightforward.

The binned energy densities $\epsilon_k^{(J)}$ can be viewed as a
step function of $x \in [0,1]$:
\begin{equation}
\label{epsx}
  \epsilon^{(J)}(x)
    =  \sum_{k=0}^{2^J-1} \epsilon_k^{(J)} \phi_{Jk}(x)
       \quad ,
\end{equation}
where
\begin{equation}
\label{boxdef}
  \phi_{Jk}(x)
    =  \phi(2^Jx-k)
    =  \left\{  \begin{array}{l}
                1  \quad\quad  {\rm for} \;
                   k 2^{-J} \leq x \leq (k+1) 2^{-J}
                   \quad ,  \\
                0  \quad\quad  {\rm else}
                \end{array}
       \right.
\end{equation}
are simple box functions.
In this respect the energy densities $\epsilon_k^{(J)}$ can be
interpreted as amplitudes of the orthogonal expansion of
$\epsilon^{(J)}(x)$ in terms of the basis functions $\phi_{Jk}(x)$ at
scale $J$.

The fundamental dilation equations
\begin{eqnarray}
\label{dileq}
  \phi_{j-1,k}(x)
    & = &  \phi(2^{j-1}x-k)
           \, = \,
           \sum_m c_m \phi(2^jx-2k-m)
           \nonumber \\
    & = &  \sum_m c_m \phi_{j,2k+m}(x)
           \quad ,  \nonumber  \\
  \psi_{j-1,k}(x)
    & = &  \psi(2^{j-1}x-k)
           \, = \,
           \sum_m (-1)^m c_{1-m} \phi(2^jx-2k-m)
           \nonumber  \\
    & = &  \sum_m (-1)^m c_{1-m} \phi_{j,2k+m}(x)
\end{eqnarray}
define a multiresolution analysis \cite{MAL89}-\cite{MEY92}. The
associated
functions $\phi$ and $\psi$ are called scaling function and wavelet,
respectively. For the box functions (\ref{boxdef}) we have
$c_0 = c_1 = 1$ and $c_k = 0$ otherwise. A different choice for
admissible coefficients $c_k$ corresponds to a different filter for the
multiresolution analysis and leads to a different wavelet $\psi(x)$;
see ref.\cite{DAU88} for a variety of examples.

In Eq.\ (\ref{dileq}) the box functions
$\phi_{Jk}(x)$ at scale $J$ are expressed in terms of the box functions
$\phi_{J-1,k^\prime}(x)$ at scale $J-1$ and the ``difference functions"
$\psi_{J-1,k^{\prime\prime}}(x)$ at scale $J-1$ with
\begin{equation}
\label{haardef}
  \psi(x)
    =  \left\{  \begin{array}{l}
                   \;\;\;
                1  \quad\quad  {\rm for} \;
                   0 \leq x < \frac{1}{2}
                   \quad ,  \\
               -1  \quad\quad  {\rm for} \;
                   \frac{1}{2} \leq x < 1
                   \quad .
               \end{array}
       \right.
\end{equation}
The set of functions $\psi_{jk}(x)$ is called Haar-wavelet basis.
Further iterating Eqs.\ (\ref{dileq}), the
energy density step function $\epsilon^{(J)}(x)$ of Eq.\ (\ref{epsx})
can be expressed as a double sum
\begin{equation}
\label{epsxwav}
  \epsilon^{(J)}(x)
    =  \sum_{j =0}^{J-1} \sum_{k =0}^{2^{j}-1}
         \tilde{\epsilon}_{jk}^{(J)}
         \psi_{jk}(x)
       + \epsilon_0^{(0)} \phi_{00}(x)
\end{equation}
representing the multiscale decomposition.
In other words, $\epsilon^{(J)}(x)$ is dissected into contributions
from different scales $j$.
Eq.\ (\ref{epsxwav}) defines the linear wavelet transformation,
which is governed by the coefficients $c_k$ entering
Eq.\ (\ref{dileq}):
\begin{equation}
\label{epstrans}
  \vec{\tilde{\epsilon}}
    =  {\bf W} \vec{\epsilon}
    =  {\bf W}(c_k) \vec{\epsilon}
\end{equation}
with
\begin{eqnarray}
\label{epsvec}
  \vec{\epsilon}
    & = &  (\epsilon_0, \, \epsilon_1, \, \ldots ,
            \epsilon_{2^J-1})
           \quad ,  \nonumber  \\
  \vec{\tilde{\epsilon}}
    & = &  (\epsilon_0^{(0)}, \, \tilde{\epsilon}_{00}, \,
            \tilde{\epsilon}_{10}, \, \tilde{\epsilon}_{11}, \,
            \tilde{\epsilon}_{20}, \, \ldots , \,
            \tilde{\epsilon}_{J-1,2^{J-1}-1})
            \nonumber  \\
    & =: &  (\tilde{\epsilon}_0, \, \tilde{\epsilon}_1, \,
             \tilde{\epsilon}_2, \, \ldots , \,
             \tilde{\epsilon}_{2^J-1})
             \quad ,
\end{eqnarray}
where the upper index $(J)$ has been omitted.
The transformed amplitudes
$\tilde{\epsilon}_{jk}^{(J)}$ are called wavelet amplitudes.
The explicit form of the
transformation matrix {\bf W} is exhibited for example in ref.\
\cite{GRE94}.

\subsection{Wavelet transformed correlation densities}

Correlations between the transformed amplitudes
$\tilde{\epsilon}_{jk}$ are called wavelet correlations and have been
introduced already in previous publications \cite{FLA92,GRE94}.
The wavelet correlation densities follow by applying the
wavelet transformation either directly
to the conventional correlation densities or
to the evolution equation for the corresponding generating function.

If the bin correlations (\ref{bincorr}) are
known, the wavelet correlations can be directly deduced with the help of
(\ref{epstrans}):
\begin{eqnarray}
\label{wacotran}
  \tilde{\rho}_{k_1}
    & \equiv &
           \left\langle \tilde{\epsilon}_{k_1} \right\rangle
           = \left\langle \sum_{k_2} {\bf W}_{k_1k_2}
             \epsilon_{k_2} \right\rangle
           =
           \sum_{k_2} {\bf W}_{k_1k_2}
           \left\langle \epsilon_{k_2} \right\rangle
           = \sum_{k_2} {\bf W}_{k_1k_2} \rho_{k_2}
           \quad ,  \nonumber  \\
  \tilde{\rho}_{k_1k_3}
    & \equiv &
           \left\langle \tilde{\epsilon}_{k_1}
                        \tilde{\epsilon}_{k_3} \right\rangle
           =
           \sum_{k_2k_4} {\bf W}_{k_1k_2} {\bf W}_{k_3k_4}
           \rho_{k_2k_4}
           \quad ,  \nonumber  \\
  \tilde{\rho}_{k_1k_3k_5}
    & \equiv &
           \left\langle \tilde{\epsilon}_{k_1}
                        \tilde{\epsilon}_{k_3}
                        \tilde{\epsilon}_{k_5} \right\rangle
           =
           \sum_{k_2k_4k_6} {\bf W}_{k_1k_2} {\bf W}_{k_3k_4}
                            {\bf W}_{k_5k_6}
           \rho_{k_2k_4k_6}
           \quad .
\end{eqnarray}
Thus the wavelet correlations can be obtained directly from the standard
bin correlations.
The treatment of wavelet correlations is general and does not
depend on the specific choice of admissible coefficients $c_k$ defining
different wavelets.

For the case of the simple Haar wavelet
it is instructive to deduce the wavelet correlations directly from
the generating function (\ref{genfunc}).
This gives more analytical insight into the correlation structure
especially for the binary
branching processes we have introduced in section 2.
In the exponent of the generating function (\ref{genfunc}) we introduce
the change into the Haar wavelet basis such that
\begin{eqnarray}
\label{etalam1}
  \sum_{k_1=0}^{2^J-1} \lambda_{k_1} \epsilon_{k_1}
    & = &  \sum_{k_1,k_3=0}^{2^J-1} \lambda_{k_1}
           \delta_{k_1k_3} \epsilon_{k_3}
           =
           \sum_{k_1,k_2,k_3=0}^{2^J-1} \lambda_{k_1}
           ({\bf W^{-1}})_{k_1k_2}  ({\bf W})_{k_2k_3}
           \epsilon_{k_3}
           \nonumber \\
    & = &  \sum_{k_1,k_2=0}^{2^J-1}
           ({\bf W^{-1}})_{k_1k_2}  \lambda_{k_1}
           \tilde{\epsilon}_{k_2}
           \equiv
           \sum_{k_2=0}^{2^J-1} \eta_{k_2} \tilde{\epsilon}_{k_2}
           \quad .
\end{eqnarray}
The coordinates $\eta_k$
show a simple relation with respect to the $\lambda_k$:
\begin{eqnarray}
\label{etalam2}
  \eta_0
    & = &  \sum_{k=0}^{2^J-1} \lambda_k
           \quad , \nonumber \\
  \eta_1
    & = &  \eta_{00} =
           \sum_{k=0}^{2^{J-1}-1} \lambda_k
           - \sum_{k=2^{J-1}}^{2^{J}-1} \lambda_k
           \quad , \nonumber \\
  \eta_2
    & = &  \eta_{10} =
           \sum_{k=0}^{2^{J-2}-1} \lambda_k
           - \sum_{k=2^{J-2}}^{2^{J-1}-1} \lambda_k
           \quad , \quad\quad\quad
           \eta_3 = \eta_{11} =
           \sum_{k=2^{J-1}}^{3 \cdot 2^{J-2}-1} \lambda_k
           - \sum_{k=3 \cdot 2^{J-2}}^{2^{J}-1} \lambda_k
           \quad , \nonumber \\
    & \vdots &  \\
  \eta_{2^{J-1}}
    & = &  \eta_{J-1,0} =
           \lambda_0 - \lambda_1
           \quad , \quad\quad \ldots , \quad
           \eta_{2^J-1} = \eta_{J-1,2^{J-1}-1} =
           \lambda_{2^J-2} - \lambda_{2^J-1}
           \quad ; \nonumber
\end{eqnarray}
for more details see again ref.\ \cite{GRE94}. In view of the
evolution equation (\ref{genevolu}) with the splitting rules
(\ref{lamevo1}) it is necessary to express
$\vec{\eta}^{(j+1)}$ at scale $j$+1 in terms of $\vec{\eta}_L^{(j)}$,
$\vec{\eta}_R^{(j)}$ and $\vec{\eta}_M^{(j)}$ at scale $j$; we deduce:
\begin{eqnarray}
\label{etaevo}
  \vec{\eta}^{(j+1)}
    & = &  \left( \eta_0^{(j+1)}, \; \eta_{00}^{(j+1)}, \;
             \eta_{10}^{(j+1)}, \; \eta_{11}^{(j+1)}, \;
             \eta_{20}^{(j+1)}, \; \ldots , \;
             \eta_{j,2^j-1}^{(j+1)}
           \right)  \quad ,  \nonumber  \\
  \vec{\eta}_L^{(j)}
    & = &  \left( \frac{1}{2} (\eta_0^{(j+1)}+\eta_{00}^{(j+1)}), \;
             \eta_{10}^{(j+1)}, \; \eta_{20}^{(j+1)}, \;
             \eta_{21}^{(j+1)}, \; \eta_{30}^{(j+1)}, \; \ldots , \;
             \eta_{j,2^{j-1}-1}^{(j+1)}
           \right)  \nonumber  \\
    & =: & \left( \eta_{0;L}^{(j)}, \; \eta_{00;L}^{(j)}, \;
             \eta_{10;L}^{(j)}, \; \eta_{11;L}^{(j)}, \;
             \eta_{20;L}^{(j)}, \; \ldots , \;
             \eta_{j-1,2^{j-1}-1;L}^{(j)}
           \right)  \quad ,  \nonumber  \\
  \vec{\eta}_R^{(j)}
    & = &  \left( \frac{1}{2} (\eta_0^{(j+1)}-\eta_{00}^{(j+1)}), \;
             \eta_{11}^{(j+1)}, \; \eta_{22}^{(j+1)}, \;
             \eta_{23}^{(j+1)}, \; \eta_{34}^{(j+1)}, \; \ldots , \;
             \eta_{j,2^{j}-1}^{(j+1)}
           \right)  \nonumber  \\
    & =: & \left( \eta_{0;R}^{(j)}, \; \eta_{00;R}^{(j)}, \;
             \eta_{10;R}^{(j)}, \; \eta_{11;R}^{(j)}, \;
             \eta_{20;R}^{(j)}, \; \ldots , \;
             \eta_{j-1,2^{j-1}-1;R}^{(j)}
           \right)  \quad ,  \nonumber  \\
  \vec{\eta}_M^{(j)}
    & = &  \left( \eta_0^{(j+1)}, \; \eta_{00}^{(j+1)}, \;
             \eta_{10}^{(j+1)}, \; \eta_{11}^{(j+1)}, \;
             \eta_{20}^{(j+1)}, \; \ldots , \;
             \eta_{j-1,2^{j-1}-1}^{(j+1)}
           \right)  \nonumber  \\
    & =: & \left( \eta_{0;M}^{(j)}, \; \eta_{00;M}^{(j)}, \;
             \eta_{10;M}^{(j)}, \; \eta_{11;M}^{(j)}, \;
             \eta_{20;M}^{(j)}, \; \ldots , \;
             \eta_{j-1,2^{j-1}-1;M}^{(j)}
           \right)  \quad .
\end{eqnarray}
This sets the stage for the evolution equation of the generating
function for the Haar wavelet correlations; from Eq.\ (\ref{genevolu})
we deduce straightforwardly
\begin{eqnarray}
\label{genevwav}
  \lefteqn{Z^{(j+1)}\!\left[\vec{\eta}^{(j+1)}\right] =}
  \nonumber  \\
    & &  (1-\tilde{p}) Z^{(j)}\!\left[\vec{\eta}_M^{(j)}\right]
         + \tilde{p} \int {\rm d}q_L {\rm d}q_R \, p(q_L,q_R) \,
           Z^{(j)}\!\left[q_L\vec{\eta}_L^{(j)}\right]
           Z^{(j)}\!\left[q_R\vec{\eta}_R^{(j)}\right]
         \quad .
\end{eqnarray}
Taking the derivatives with respect to the $\eta_{j_1k_1}^{(j+1)}$,
recursion relations for the Haar wavelet correlations can be found. For
the first order Haar wavelet correlations we find
\begin{eqnarray}
\label{corwav1}
  \tilde{\rho}_0^{(j+1)}
    & = &  \frac{1}{i} \left.
           \frac{\partial Z^{(j+1)}[\vec{\eta}^{(j+1)}]}
                {\partial \eta_0^{(j+1)}}
           \right|_{\vec{\eta}^{(j+1)}=0} =
           (1-\tilde{p} ) \tilde{\rho}_0^{(j)}
           + \tilde{p} \frac{\overline{q_L}+\overline{q_R}}{2}
           \tilde{\rho}_0^{(j)}
           \quad ,  \nonumber  \\
  \tilde{\rho}_{(00)}^{(j+1)}
    & = &  (1-\tilde{p}) (1-\delta_{j0}) \tilde{\rho}_{(00)}^{(j)}
           + \tilde{p} \frac{\overline{q_L}-\overline{q_R}}{2}
           \tilde{\rho}_0^{(j)}
           \quad ,  \nonumber  \\
  \tilde{\rho}_{(j_1k_1)}^{(j+1)}
    & = &  (1-\tilde{p}) (1-\delta_{jj_1})
           \tilde{\rho}_{(j_1k_1)}^{(j)}
           + \tilde{p} \, \overline{q} \,
           \tilde{\rho}_{(j_1-1,k_2)}^{(j)}
           \quad ;
\end{eqnarray}
the last relation holds for $1 \leq j_1 \leq j$ and
$\overline{q}=\overline{q_L}$, $k_2=k_1$ for $(j_1k_1) \in \{ L \}$ or
$\overline{q}=\overline{q_R}$, $k_2=k_1-2^{j_1-1}$
for $(j_1k_1) \in \{ R \}$. The starting value is
$\tilde{\rho}_0^{(0)} = \rho_0^{(0)} = 1$. Furthermore, we specify three
recursion relations for the second order Haar wavelet correlations,
which will turn out to be important for later discussions:
\begin{eqnarray}
\label{corwav2}
  \tilde{\rho}_{0,0}^{(j+1)}
    & = &  \frac{1}{i^2} \left.
           \frac{\partial^2 Z^{(j+1)}[\vec{\eta}^{(j+1)}]}
                {\partial (\eta_0^{(j+1)})^2}
           \right|_{\vec{\eta}^{(j+1)}=0}
           \nonumber  \\
    & = &  (1-\tilde{p}) \tilde{\rho}_{0,0}^{(j)}
           + \tilde{p} \frac{(\overline{q_L^2}+\overline{q_R^2})}{4}
           \tilde{\rho}_{0,0}^{(j)}
           + \tilde{p} \frac{\overline{q_Lq_R}}{2}
           (\tilde{\rho}_{0}^{(j)})^2
           \quad ,  \nonumber  \\
  \tilde{\rho}_{(00),(00)}^{(j+1)}
    & = &  (1-\tilde{p}) (1-\delta_{j0}) \tilde{\rho}_{(00),(00)}^{(j)}
           + \tilde{p} \frac{(\overline{q_L^2}+\overline{q_R^2})}{4}
           \tilde{\rho}_{0,0}^{(j)}
           - \tilde{p} \frac{\overline{q_Lq_R}}{2}
           (\tilde{\rho}_{0}^{(j)})^2
           \quad ,  \nonumber  \\
  \tilde{\rho}_{(j_1k_1),(j_1k_1)}^{(j+1)}
    & = &  (1-\tilde{p}) (1-\delta_{jj_1})
           \tilde{\rho}_{(j_1k_1),(j_1k_1)}^{(j)}
           + \tilde{p} \, \overline{q^2} \,
           \tilde{\rho}_{(j_1-1,k_2),(j_1-1,k_2)}^{(j)}
           \quad ,
\end{eqnarray}
where again $\overline{q^2}=\overline{q_L^2}$ or
$\overline{q_R^2}$ and $k_2=k_1$ or $k_1-2^{j_1-1}$ for
$(j_1k_1) \in \{ L \}$ or $\{ R \}$ in the last relation.

\subsection{Wavelet correlations of specific cascade models}

\subsubsection{$p$-model}

Wavelet correlations in the $p$-model were already studied in ref.\
\cite{GRE94}, emphasizing the selfsimilarity aspect of the wavelet
basis. We briefly recall the main results.
Since the splitting moments $\overline{q_L}$ and $\overline{q_R}$ are
equal (see Eq.\ (\ref{splitmom})), the first order Haar wavelet
correlations (\ref{corwav1}) are all zero,
\begin{equation}
\label{corwav1o}
  \tilde{\rho}_{(jk)} = 0  \quad ,
\end{equation}
except for $\tilde{\rho}_0=1$.
On average a difference between energy densities of
neighbouring bins is zero, as a difference might come with a positive or
negative sign with equal probability. Only the global average
$\tilde{\rho}_0$ of all bins is one.

{}From (\ref{corwav2}) and (\ref{splitmom}) we deduce for the second order
Haar wavelet correlations
\begin{eqnarray}
\label{corwav2o}
  \tilde{\rho}_{0,0}
    & = &  1   \quad ,  \nonumber  \\
  \tilde{\rho}_{(j_1k_1),(j_2k_2)}
    & = &  \alpha^2 (1+\alpha^2)^{j_1} \delta_{j_1j_2} \delta_{k_1k_2}
           \quad ;
\end{eqnarray}
they are depicted in Fig.\ 6a, where the wavelet indices $(jk)$ are
ordered according to Eq.~(\ref{epsvec}).
The product of two differences belonging to different scales or bin
positions is zero on average. The diagonal contributions, which
represent squares of differences, show a scaling law as the resolution
$j_1$ increases; compare with (\ref{momp}).
It is exactly this diagonal structure we would have
expected from second order Haar wavelet correlations of the selfsimilar
$p$-model cascade.
In ref.\ \cite{GRE94} we have shown that when
choosing wavelets other than the Haar wavelet, the second order wavelet
correlations of the $p$-model cascade turn out to be quasidiagonal,
with diagonal contributions dominating over the off-diagonal ones and
also showing an approximate scaling law.

For the Haar wavelet correlations of the third order a
double scaling is found:
\begin{equation}
\label{doubscal}
  \tilde{\rho}_{(j_1k_1),(j_2k_2),(j_2k_2)}
    =  (1+3\alpha^2)^{j_1}
       \left( \pm 2\alpha^4(1+\alpha^2)^{j_2-j_1-1} \right)
       \quad ,
\end{equation}
where $(j_1k_1)$, $(j_2k_2)$ have to share a common parenthood,
i.e.\ $0 \leq j_1 < j_2$,
$k_1 2^{j_2-j_1} \leq k_2 \leq k_1 2^{j_2-j_1} + 2^{j_2-j_1-1} -1$
for the (+) sign and
$k_1 2^{j_2-j_1} + 2^{j_2-j_1-1} \leq k_2 \leq (k_1+1) 2^{j_2-j_1} -1$
for the (--) sign. All the other pure third order Haar wavelet
correlations
are zero; only some of those involving an index 0, which represents the
amplitude $\tilde{\epsilon}_0=1$ of the global average and thus not a
difference amplitude, are nonzero but would vanish for
Haar wavelet cumulants \cite{GRE94}.

Note that the wavelet transform compresses the full information
contained in the second order correlation function (\ref{rhobin2}) into
the diagonal (\ref{corwav2o}); compare also Fig.~4a with Fig.~6a. In
addition to the diagonal contributions also certain off-diagonal
bandstructures arise in higher order wavelet correlations as e.g.\ in
Eq.~(\ref{doubscal}). An interpretation of these offdiagonal
contributions in terms of clump correlations is given in section 3.4.

\subsubsection{$\alpha$-model}

{}From Eqs.\ (\ref{corwav1}) and
(\ref{splitmoma}) we derive exactly the same result for the first order
Haar wavelet correlations of the $\alpha$-model as stated in eq.\
(\ref{corwav1o}) for the $p$-model. Differences between the two arise
for
higher order Haar wavelet correlations. With the branching probability
$\tilde{p}=1$ and the second order splitting moments of
(\ref{splitmoma}), we derive
\begin{eqnarray}
\label{alpwav2}
  \tilde{\rho}_{0,0}^{(J)}
    & = &  \frac{1}{(1-\alpha\beta)} \left\{
           1 - \alpha\beta\left( \frac{1+\alpha\beta}{2} \right)^J
           \right\}  \quad ,  \nonumber  \\
  \tilde{\rho}_{(j_1k_1),(j_2k_2)}^{(J)}
    & = &  \frac{\alpha\beta (1+\alpha\beta)^{j_1}}{(1-\alpha\beta)}
           \left\{  1 - \left( \frac{1+\alpha\beta}{2} \right)^{J-j_1}
           \right\}  \delta_{j_1j_2} \delta_{k_1k_2}
\end{eqnarray}
from the recursion relations (\ref{corwav2}) for the second order Haar
wavelet correlations.
This correlation matrix, which is depicted in
Fig.\ 6b, is diagonal as in the $p$-model case. However,
the diagonal contributions now depend on the
number of performed cascade steps $J$.

We first comment on the element
$\tilde{\rho}_{0,0}^{(J)}$. In contrast to the $p$-model result,
it is unequal to one; for
$| \alpha\beta | < 1$ it is equal to $1 / (1-\alpha\beta)$ in the limit
$J \rightarrow \infty$. The quantity
$\tilde{\rho}_{0,0}^{(J)} - 1$ reflects the square width of the
fluctuation in the total energy. Its deviation
from zero is a clear evidence that the total energy is not conserved in
the $\alpha$-model cascade.

The elements
$\tilde{\rho}_{(j_1k_1),(j_1k_1)}^{(J)}$, reflecting the true second
order Haar wavelet correlations, do not show a rigorous scaling; only
for $j_1$ values being ``safely" smaller than the number of cascade
steps
$J$ and $| \alpha\beta | < 1$ does a power law depending on the scale
index
$j_1$ show up. If we set $\alpha = \beta$ in the $\alpha$-model and use
the same value as in the $p$-model, we obtain the same asymptotic
scaling as for the perfect scaling in the
$p$-model. The only difference is that the missing anticorrelations
lead to a modification of $1 / (1-\alpha\beta)$ in the overall factor.

In Fig.\ 4, we show the deviation from
scaling of the diagonal elements of the
second order Haar wavelet transformed correlations
$\tilde{\rho}_{(jk),(jk)}^{(J)}$ with respect to the scale index
$j$ for both the $\alpha$- and the $p$-model.
Note the resemblance between Eq.~(\ref{alpwav2}) and the moment
Eq.~(\ref{moma}), which emphasizes again that the wavelet correlations
imply, by design, a backward analysis. Note also by comparison of
Figs.~3b and 6b that these deviations from perfect scaling are
observed more easily in the wavelet picture.

\subsubsection{$p$-model with random branching}

We do not learn anything new from the recursion relations
(\ref{corwav1}) for the first order Haar wavelet correlations.
The global
average is $\tilde{\rho}_0^{(J)} = 1$; first order differences vanish
for any $\tilde{p}$, so that $\tilde{\rho}_{(jk)}^{(J)} = 0$.

Turning to the more interesting second order Haar wavelet
correlations (\ref{corwav2}), we
find, as expected, that $\tilde{\rho}_{0,0}^{(J)} = 1$.
Again, energy is conserved in every possible configuration. Off-diagonal
contributions to the second order Haar wavelet correlation matrix
$\tilde{\rho}_{(j_1k_1),(j_2k_2)}^{(J)}$ vanish also in this more
general cascade.
Next, we consider the two diagonal elements
$\tilde{\rho}_{(0,0),(0,0)}^{(J)}$ and
$\tilde{\rho}_{(J-1,k),(J-1,k)}^{(J)}$
in particular.

For $\tilde{\rho}_{(0,0),(0,0)}^{(J)}$ we find
\begin{equation}
\label{pmrb00}
  \tilde{\rho}_{(0,0),(0,0)}^{(J)}
    =  \alpha^2 \left( 1 - (1-\tilde{p})^J \right)
       \quad .
\end{equation}
The factor
$(1-\tilde{p})^J$ represents the probability that after $J$ cascade
steps no branching has occurred. For $0 < \tilde{p} < 1$ and
$J \rightarrow \infty$ this probability goes to zero. Then
$\tilde{\rho}_{(0,0),(0,0)}^{(J)}$, which represents the square of the
difference between the average of the energy densities belonging to the
$2^{J-1}$ left bins and the average of those belonging to the
$2^{J-1}$ right bins, approaches the value (\ref{corwav2o})
obtained for the $p$-model with $\tilde{p}=1$.

The element $\tilde{\rho}_{(J-1,k),(J-1,k)}^{(J)}$
representing the square of the difference
between energy densities in two adjacent bins at the finest scale
follows from the third equation of (\ref{corwav2}):
\begin{equation}
\label{pmrbjj}
  \tilde{\rho}_{(J-1,k),(J-1,k)}^{(J)}
    =  \tilde{p} \alpha^2
       \left( \tilde{p} (1+\alpha^2) \right)^{J-1}
       \quad .
\end{equation}
For $\tilde{p} < 1/(1+\alpha^2)$ this element tends to zero as
$J \rightarrow \infty$. As the number of cascade steps increases, the
probability that no branching has occurred on the finest scales also
increases, so that the difference amplitudes
on the finest scales will be zero. Once
$\tilde{p} \geq 1/(1+\alpha^2)$ however, this interpretation no
longer holds. For $\tilde{p} = 1/(1+\alpha^2)$ we have always
$\tilde{\rho}_{(J-1,k),(J-1,k)}^{(J)} = \tilde{p} \alpha^2$, no matter
how large the number of cascade steps $J$. Somehow
this seems to contradict common sense, especially when the number of
cascade steps becomes very large. It is true that, as
$J \rightarrow \infty$, the probability of a branching occurring at
every cascade step
becomes infinitesimal small; but on the other hand,
once it happens, the difference amplitudes
on the finest scales become very large
because neighbouring bins have gone through a long common history and
have built up huge energy densities. Differences of large densities are
also large since the splitting in the very last cascade step always goes
with a $(1+\alpha)$ to the left and a $(1-\alpha)$ to the right or vice
versa. This huge difference then dominates over the small probability
that branching occurs; as a consequence the correlation
$\tilde{\rho}_{(J-1,k),(J-1,k)}^{(J)}$ stays finite and unequal zero for
$\tilde{p} = 1/(1+\alpha^2)$. For $\tilde{p} > 1/(1+\alpha^2)$ it even
increases as the number of cascade steps $J$ goes to infinity.

Fig.~6c shows the full second order Haar wavelet
correlation density for the $p$-model
with random branching for the case $\tilde{p} < 1/(1+\alpha^2)$. In
Fig.~7, the scale dependence of $\tilde{\rho}_{(j,k),(j,k)}^{(J)}$ is
shown for various values of the branching probability $\tilde{p}$. For
$j$ much smaller than $J$ the same scaling shows up for the $p$-model
with random branching as with deterministic branching; large-scale
branchings have occurred during the cascade with a probability almost
equal to one. For small branching probabilities, i.e.\
$\tilde{p} < 1/(1+\alpha^2)$, a deviation from scaling sets in at
midscales and the contributions $\tilde{\rho}_{(j,k),(j,k)}^{(J)}$ drop
rapidly to zero as finer and finer scales are considered. This is
clearly a lifetime effect, because branchings at the finest
scales will hardly occur.

For higher order Haar wavelet correlations similar modifications are to
be expected. The multiple scalings (\ref{doubscal}) and
(\ref{corwav4o}) for higher
order Haar wavelet correlations will only live on the larger
scales and will break down at the finest scales.

\subsubsection{QCD-motivated cascade model}

The first order Haar wavelet correlation densities for the QCD-motivated
cascade model follow from the recursion relation (\ref{corwav1}) using
the splitting moments (\ref{qcdspmom}) and $\tilde{p}=1$:
\begin{eqnarray}
\label{qcdwav1}
  \tilde{\rho}_0^{(J)}
    & = &  (1 + \gamma)^J
           \quad ,  \nonumber  \\
  \tilde{\rho}_{(jk)}^{(J)}
    & = &  0
           \quad .
\end{eqnarray}
As expected, the global average multiplicity density
$\tilde{\rho}_0^{(J)}$ is equal to the local average density
$\rho_k^{(J)}$ of Eq.~(\ref{qcdcor1}), whereas all local differences are
zero on average.

{}From (\ref{corwav2}) and (\ref{qcdspmom}) we deduce for the second order
Haar wavelet correlations
\begin{eqnarray}
\label{qcdwav2}
  \tilde{\rho}_{0,0}^{(J)}
    & = &  \frac{\left( 1 + \gamma \right)^{2J}}
                {\left( \frac{3}{2} + \gamma \right)}
           +
           \frac{\left( \frac{1}{2} + \gamma \right)}
                {\left( \frac{3}{2} + \gamma \right)}
           \left( 1 + \frac{\gamma}{2} \right)^J
           \quad ,  \nonumber  \\
  \tilde{\rho}_{(j_1k_1),(j_2k_2)}^{(J)}
    & = &  \left[
           \frac{\left( 1 - \gamma - \gamma^2 \right)}
                {\left( \frac{3}{2} + \gamma \right)}
           (1 + \gamma)^{2J-2}
           \left(
           \frac{2 + \gamma}{(1 + \gamma)^2}
           \right)^j
           \right.  \\
    &   &  \; \left. + \;
           \frac{\left( \frac{1}{2} + \gamma \right)}
                {\left( \frac{3}{2} + \gamma \right)}
           \left( 1 + \frac{\gamma}{2} \right)^J 2^j
           \right]
           \delta_{j_1j_2} \delta_{k_1k_2}
           \quad .  \nonumber
\end{eqnarray}
The average square of the total multiplicity density
$\tilde{\rho}_{0,0}^{(J)}$ is unequal to
$(\tilde{\rho}_0^{(J)})^2$, signaling the absence of proper
anticorrelations. The true second order Haar wavelet correlations
$\tilde{\rho}_{(j_1k_1),(j_2k_2)}^{(J)}$ are again diagonal; normalized
with respect to $(\tilde{\rho}_0^{(J)})^2$ they are shown in Fig.~6d.
The diagonal contributions depend on the number of performed cascade
steps $J$ and do not show rigorous scaling. In fact, as in the case of
(backward) moments (\ref{qcdm2}), the dependence on the resolution scale
$j$ splits into a term with the same anomalous scaling and another term
scaling trivially in powers of 2. This $j$-dependence is once
more illustrated in Fig.~5; the deviation from the anomalous scaling at
the rougher resolution scales sets in earlier for the wavelet
correlations than for the (backward) moments.

It is instructive to introduce wavelet moments $W_q(J,j)$. For the case
of second order we define
\begin{equation}
\label{qcdwavm}
  W_2(J,j)
    =  \frac{1}{2^j} \sum_{k=0}^{2^j-1}
       \tilde{\rho}_{(jk),(jk)}^{(J)}
       \; = \;
       \tilde{\rho}_{(j0),(j0)}^{(J)}
       \quad .
\end{equation}
They are related to the (backward) moments $M_2(J,j)$ of (\ref{mom3})
via
\begin{eqnarray}
\label{qcdmr}
  W_2(J,J-j-1)
    & = &  M_2(J,J-j) - M_2(J,J-j-1)
           \nonumber  \\
    & = &  \frac{1}{2} \left(
           M_2(J,J-j) - \rho_{0,2^j}^{(J)}
           \right)  \quad ,
\end{eqnarray}
which follows from Eq.~(\ref{wacotran}).
This relation reflects the multiresolution property of the wavelets as
they look on the difference of two adjacent scales. A ``difference"
moment is more sensitive to deviations from perfect scaling than an
``average" moment; this explains the results depicted in Fig.~5. Also,
in view of the relation (\ref{qcdmr}), it now becomes clear that the
expression (\ref{qcdwav2}) for the wavelet correlations splits into the
same two different scaling terms as the (backward)
moments (\ref{qcdm2}) did.

Expressions for higher order wavelet correlation densities will not be
given here for the QCD-motivated cascade model. Their general
interpretation will be given in the next two sections 3.4 and 3.5. For
a gluon cascade they can be understood as correlations between gluon
subjets inside larger gluon jets.

\subsection{Higher order wavelet correlations: clumps}

What does the double scaling of the third order wavelet correlation
(\ref{doubscal}) tell us? To get a glimpse, we
consider once again the evolution of the energy
densities according to the cascading prescription of the $p$-model.
See Fig.\ 8. After $j_1$ cascade steps we pick
out the energy density $\epsilon_{k_1}^{(j_1)}$ of one bin with label
$k_1$.
At the next cascade step this bin has split into two subbins with
labels $2k_1$ and $2k_1+1$ respectively, which contain the energy
densities
$\epsilon_{2k_1}^{(j_1+1)}$ and $\epsilon_{2k_1+1}^{(j_1+1)}$.
We ask now for the correlation
between the energy density $\epsilon_{k_2}^{(j_2)}$ contained in a
subbin with labels $j_2,k_2$ and the energy density
$\epsilon_{k_1}^{(j_1)}$
contained in the picked bin with labels $j_1,k_1$
where the support of the
subbin $j_2,k_2$ is contained in the support of the bin $j_1,k_1$; i.e.\
$j_1 < j_2$ and
$k_1 2^{j_2-j_1} \leq k_2 \leq (k_1+1) 2^{j_2-j_1} -1$. We get for
example:
\begin{eqnarray}
\label{cluscor}
  \left\langle \epsilon_{k_1}^{(j_1)} \epsilon_{k_2}^{(j_2)}
  \right\rangle
    & = &  \left\langle ( \epsilon_{k_1}^{(j_1)} )^2 \right\rangle
           \left\langle \epsilon_{k_3}^{(j_2-j_1)} \right\rangle
           =  (1+\alpha^2)^{j_1}
           \quad ,  \nonumber  \\
  \left\langle \epsilon_{k_1}^{(j_1)}
  ( \epsilon_{k_2}^{(j_2)} )^2 \right\rangle
    & = &  \left\langle ( \epsilon_{k_1}^{(j_1)} )^3 \right\rangle
           \left\langle ( \epsilon_{k_3}^{(j_2-j_1)} )^2 \right\rangle
           =  (1+3\alpha^2)^{j_1} (1+\alpha^2)^{j_2-j_1}
           \quad ,
\end{eqnarray}
where $k_3 = k_2-k_1 2^{j_2-j_1}$. These interscale correlations contain
additional
information about the subclustering structure of the cascade process.
In other words, this can be interpreted as the correlation between a
``coherent structure" and one of its ``substructures". Such
(sub)structures can be identified with clusters (high density regions)
as well as voids (low density regions) and will subsequently be referred
to as ``clumps".

The second order correlation
$\langle \epsilon_{k_1}^{(j_1)} \epsilon_{k_2}^{(j_2)} \rangle$
only provides
information about the common ``history" of the clumps $j_1k_1$ and
$j_2k_2$ and not about the subclustering structure, because
$\langle \epsilon_{k_3}^{(j_2-j_1)} \rangle = 1$. This situation changes
once we consider the third order correlation
$\langle \epsilon_{k_1}^{(j_1)} ( \epsilon_{k_2}^{(j_2)} )^2 \rangle$.
Here we
get a double scaling, which depends on the common ``history" of the two
clumps as well as on the subclustering structure of the larger
clump.
It is the same double scaling we have found for the third order
Haar wavelet correlations (\ref{doubscal}).
Therefore we interpret higher
order Haar wavelet correlations as clump correlations. In fact, what
the third order Haar wavelet correlations do, is that
they correlate the difference of the energy densities of two
adjacent clumps living on the scale $j_1$ with the square of the
difference of the energy densities of two neighbouring subclumps
living on the scale $j_2$ inside the original clump, as depicted in
Fig.~8 by marked arrows. Of course, this is just another definition
of clump correlations.

For completeness we state the nonvanishing contributions to the fourth
order Haar wavelet correlations of the $p$-model which provide further
information about the subclumping structure:
\begin{eqnarray}
\label{corwav4o}
  \tilde{\rho}_{(j_1k_1)^4}
    & = &  \alpha^4 (1+6\alpha^2+\alpha^4)^{j_1}
           \; ,  \nonumber  \\
  \tilde{\rho}_{(j_1k_1)^2,(j_2k_2)^2}
    & = &  \alpha^4 (1+\alpha^2)^{j_1-j_2}
           (1+6\alpha^2+\alpha^4)^{j_2}
           \; ,  \nonumber  \\
    &   &  \quad\quad\quad\quad\quad\quad\quad\quad
           ( j_2<j_1, \;
           {\rm same \; parenthood} ),
           \nonumber  \\
    & = &  \alpha^4 (1-\alpha^2)^2
           (1+6\alpha^2+\alpha^4)^{s}
           (1+\alpha^2)^{j_1+j_2-2s-2}
           \; ,  \nonumber  \\
    &   &  \quad\quad\quad\quad\quad\quad\quad\quad
           ( {\rm partially \; same \; parenthood} ),
           \nonumber  \\
  \tilde{\rho}_{(j_1k_1)^2,(j_2k_2),(j_3k_3)}
    & = &  \left( \pm \right)_{j_3}
           \left( \pm \right)_{j_2}
           2\alpha^6 (3+\alpha^2) (1+6\alpha^2+\alpha^4)^{j_3}
           (1+3\alpha^2)^{j_2-j_3-1} \cdot
           \nonumber  \\
    &   &  \cdot (1+\alpha^2)^{j_1-j_2-1}
           \; , \nonumber \\
    &   &  \quad\quad\quad\quad\quad\quad\quad\quad
           ( j_3<j_2<j_1, \;
           {\rm same \; parenthood} ).
\end{eqnarray}
The phrase ``same parenthood" used in the first case of
$\tilde{\rho}_{(j_1k_1)^2,(j_2k_2)^2}$ stands for
$k_2 2^{j_1-j_2} \leq k_1 \leq (k_2+1) 2^{j_1-j_2}-1$. The second case
with ``partially same par\-ent\-hood" translates into $0 \leq k \leq
2^s-1$,
$\, k \cdot 2^{j_1-s} \leq k_1 \leq (k+\frac{1}{2}) 2^{j_1-s}-1$,
$\, (k+\frac{1}{2}) 2^{j_2-s} \leq k_2 \leq (k+1)2^{j_2-s}-1$,
where $s$ represents the scale, from which point the wavelet indices
$(j_1k_1)$ and $(j_2k_2)$ follow different branches in the underlying
tree structure.
($j_1-s$) and ($j_2-s$) are the scales of two different subclumps
relative to the scale $s$ of their common parent clump. Thus
the correlation density $\tilde{\rho}_{(j_1k_1)^2,(j_2k_2)^2}$
provides information not only about the direct subclustering of a
large clump,
but also relates two different subclumps within a large clump.

Furthermore, the triple scaling of
$\tilde{\rho}_{(j_1k_1)^2,(j_2k_2),(j_3k_3)}$
tells us something about the
correlation of a subsubclump within a subclump within a clump. Here the
phrase "same parenthood" illustrates
$k_3 \cdot 2^{j_2-j_3} \leq k_2 \leq (k_3+\frac{1}{2}) 2^{j_2-j_3}-1$
for the (+) sign and
$(k_3+\frac{1}{2}) 2^{j_2-j_3} \leq k_2 \leq (k_3+1) 2^{j_2-j_3}-1$
for the (--) sign in $(\pm)_{j_3}$ and, furthermore,
$k_2 \cdot 2^{j_1-j_2} \leq k_1 \leq (k_2+\frac{1}{2}) 2^{j_1-j_2}-1$
for the (+) sign and
$(k_2+\frac{1}{2}) 2^{j_1-j_2} \leq k_1 \leq (k_2+1) 2^{j_1-j_2}-1$
for the (--) sign in $(\pm)_{j_2}$.
-- This demonstrates once more that higher order wavelet correlations
provide direct information about the clustering hierarchy and can be
interpreted as clump correlations. In the next section we further
illuminate this interpretation from a two-dimensional perspective.

We emphasize that this kind of subclustering information cannot be
obtained by a conventional moment analysis along the lines of section
2.4. Although in principle this information is contained in the
conventional higher-order correlation densities,
it is very hard to measure or visualize directly. Due to its
multiresolution character, the wavelet transform compresses exactly this
information of interest to a readily accessible form.

\subsection{Multiscale clustering \newline
            in two-dimensional branching processes}

To illustrate and visualize clumps in more detail, we will now
consider hierarchical branching processes in two dimensions. As a
representative we choose the two-dimensional $\alpha$-model. Compared to
the $p$-model, the $\alpha$-model is easier to generalize to higher
dimensions, as it can be built by direct products of  one-dimensional
$\alpha$-models. Instead of halving a given interval, a
square is subdivided into four subsquares with labels 1,\ldots,4
during one cascade step.
The original energy is partitioned according to the splitting function
\begin{equation}
\label{clump1}
  p(q_1,\ldots,q_4)
    =  \prod_{i=1}^{4} \big[
       p_1 \delta(q_i-(1-\alpha)) + p_2 \delta(q_i-(1+\beta))
       \big]
       \quad ;
\end{equation}
this is a straightforward generalisation of eq.\ (\ref{splitalp}).
The prescription (\ref{clump1}) is repeated for all following cascade
steps. One possible realization of the two-dimensional $\alpha$-model
is depicted in the upper left corner of
Fig.~9; six cascade steps ($j=6$)
have been performed with the splitting parameters $\alpha = \beta = 0.4$
We have used a gray scale to indicate the population of
regions in between large energy densities (white) and small energy
densities (black). We observe that certain regions clump into clusters
of large densities and other regions with small energy densities exist
as voids. Furthermore clusters/voids appear in different sizes and show
substructures, which are again clumped into clusters and voids.

To quantify this picture, we explicitly perform
a multiresolution analysis: First
the energy densities of four little squares constituting a larger
square are averaged. These averaged energy densities are depicted in
Fig.~9 as the second figure from the top of the left column. In other
words, the original configuration has been smoothed or resolved on a
rougher scale $j$=5. Some detail is obviously lost in this
representation; no information about substructures of clumps living on
the scale $j=5$ can be deduced. This lost information can be recovered
by keeping the difference between the resolutions $j$=5 and $j$=6;
this difference information is illustrated in the picture
to the right of the $j$=5 resolution picture. If
this difference vanishes in some regions, no substructures are present;
if on the other side those differences become large in other regions,
then this indicates sizeable substructures on the scale $j$=6.

These smoothing and differentiation operations are iterated through
scales $j$=4, 3 down to 2.
In general, the difference between two successive smoothings
of a configuration gives the information about the subclustering present
at the involved scales. It is important to notice that the difference
information on one scale is completely independent of
(orthogonal to) the difference
information on any other scale and therefore does not carry along
redundant information. Also, no information is lost by
keeping only the differences of smoothed representations. The sum of
all difference representations together with the roughest smoothed
configuration ($j$=2) recovers
the original configuration belonging to  the finest resolution scale.
This is the two-dimensional analogue to Eq.~(\ref{epsxwav}).

The smoothing operations at scale $j$ are performed with the basis
\begin{equation}
\label{clump2}
  \phi_{j,k_1k_2}(x,y)
    =  \phi(2^{-j}x-k_1,2^{-j}y-k_2)
    =  \phi(2^{-j}x-k_1) \phi(2^{-j}y-k_2)
    =  \phi_{jk_1}(x) \phi_{jk_2}(y)
       \quad ;
\end{equation}
this is a direct product of two one-dimensional scaling functions
given in eq.\ (\ref{boxdef}) as box functions. The difference
between two successive smoothing operations can be completely expressed
in terms of two-dimensional wavelets, namely
\begin{eqnarray}
\label{clump3}
  \psi^{(1)}_{j,k_1k_2}(x,y)
    & = &  \phi_{jk_1}(x) \psi_{jk_2}(y)
           \quad ,   \nonumber  \\
  \psi^{(2)}_{j,k_1k_2}(x,y)
    & = &  \psi_{jk_1}(x) \phi_{jk_2}(y)
           \quad ,   \nonumber  \\
  \psi^{(3)}_{j,k_1k_2}(x,y)
    & = &  \psi_{jk_1}(x) \psi_{jk_2}(y)
           \quad .
\end{eqnarray}
Here $\psi_{jk}$ represents the one-dimensional Haar wavelet
(\ref{haardef}). Once again the dilation equations (\ref{dileq}) govern
the smoothing and differentiation transformations of the original
configuration. -- Referring to the
previous paragraph, the difference information,
i.e.\ the right column of Fig.~9, can be completely
expanded in the two-dimensional Haar-wavelet basis (\ref{clump3}). The
correlations between the amplitudes of this multiresolution
expansion are the wavelet correlations. Thus it becomes obvious that the
wavelet correlations do characterize the subclustering of clumps.

This interpretation works not only for the Haar wavelet basis.
In Fig.~10 a two-dimensional wavelet analysis has been
performed with the so-called Daubechies D12 wavelet \cite{DAU88}. The
decompositions (\ref{clump2}) and (\ref{clump3}) still hold; the only
difference to the Haar-wavelet analysis is that 12 coefficients
$c_m$ are used in the dilation equations (\ref{dileq}). Again the left
column of Fig.~10 shows density plots of a sequence of D12-smoothed
approximations to the original configuration, whereas the middle column
represents density plots of the wavelet transform, which form a
sequence of mutually orthogonal details. In order to provide a better
picture of the subclustering aspect of the wavelet
transform, the details at various scales are exhibited
again in the right column, but with only two gray-values: black for
regions where the detail function becomes negative, indicating local
voids, and white for regions where the
detail function is positive, signaling the appearance of
clusters at the various scales.

The borders between white and black regions in the right column of
Fig.~10 are the zero-crossings of the wavelet transform.
In signal analysis, these
zero-crossings localize the signal sharp variation
points at different scales and act as edge detectors \cite{MAL91}. The
related wavelet transform maxima method \cite{MAL92} is used to
compress and reconstruct signals efficiently.

\section{Conclusions and outlook}

For the complex dynamics of multiparticle processes we study
the fundamental differential correlation densities in order to extract
as much important statistical information (scaling, clustering) as
possible. This has  been demonstrated with the help of some simple
discrete hierarchical branching models. An evolution equation for the
(multivariate) generating function has been formulated, from which the
correlation densities follow recursively. An integration over the
correlation densities yields the moments.

Contrary to common practice,
a careful conceptual distinction has to be made between forward
(= theoretical) moments, which are obtained in evolution direction of
the cascade process, and backward (= experimental) moments
obtained in a backward analysis after the last cascade step has been
performed. The former, often calculated from theory, can not be directly
compared to the latter. Only the backward moments are accessible to
experimental observation.
We show that hierarchical branching processes with
global density fluctuations on top of local density fluctuations do not
show a rigorous scaling in the observed moments, although they are
constructed by a purely selfsimilar iteration law.
{\em In other words, a
deviation from scaling of the observed moments
does not necessarily imply that the cascade process is not
selfsimilar}.
This poses the provocative question: What is now the
essence of a moment (multifractal) analysis which only focuses on
scaling indices? A deviation from perfect scaling provides important
information on the cascade mechanism as we have demonstrated by various
models. -- In contrast to a moment
analysis, the selfsimilar (forward) scaling can still be extracted from
the differential correlation densities.

However conventional correlation densities, especially higher orders,
are hard to quantify. For their representation a compression is
needed which reveals only the important (statistical) information and
removes redundancy. In this respect the wavelet transformation
appears very appealing. Wavelets constitute a selfsimilar and
orthogonal basis; in addition, due to their multiresolution properties,
they dissect structures into details (clumps) living on different
scales. As a
consequence the wavelet transformed correlation densities of all orders
become very sparse for hierarchically organized processes.
The second order correlations diagonalize completely
in the Haar-wavelet basis. Furthermore, for any order of
the wavelet correlations, the diagonal elements show scaling exponents
which can be related to the multifractal dimensions. Beyond the
multifractal analysis, those few off-diagonal elements of the higher
order Haar-wavelet correlations which are nonzero and arise in bands
show multiple
scaling and provide information on the subclustering structure of clumps
such as correlations between small clusters living within larger
clusters or voids.

This paper provides a demonstration
of the wavelet capabilities for correlation studies.
However, the techniques presented here have to be further
refined in order to become a powerful tool for experimental analysis.
In this respect wavelet packets \cite{MEY92,RUS92}
might help to simplify the
correlation structure even further. Also, in view of the recent success
of correlation integrals \cite{EGG93}, one should develop an analogue,
so-called wavelet correlation integrals; this could set new standards
of inferring information on clump correlations.
Mathematically, there is the challenging problem to define the
wavelet transformation on spheres which might have
important applications in the context of angular intermittency.

We believe that a true understanding of the
intermittency phenomenon in multiparticle cascading is still a long way
off. For general
particle cascades it is not a priori clear that they follow a
hierarchical evolution structure.
Particle cascades should be formulated as point processes, which need
not be hierarchically organized from the beginning. The relevant
splitting
functions should dictate if hierarchical structures could evolve
dynamically. This might be the case for perturbative QCD. Work in this
direction, including the use of the wavelet transformation, is presently
being carried out.

Apart from multiparticle physics, the above outlined
wavelet technology might have
applications in other fields of physics. For example,
wavelet (clump) correlations may shed further light on the long-standing
question in intermittent fully developed turbulence of how eddies
decay at small scales and would allow to gain new and supplementary
insight for phenomenological modeling. In disordered solid state systems
multifractality and localization of electron wavefunctions are
discussed; a wavelet analysis appears to be suitable \cite{KAN95}.
Last not least,
the simplifying aspect of wavelet correlations might ignite the
century-old
inverse problem of how to extract dynamics from correlations with
new fuel.

\vspace{1cm}
\newpage

{\noindent\bf Acknowledgments:}
We are indebted to Werner Scheid, Brigitte Buschbeck and Hans Eggers
for many fruitful discussions.
M.G.\ wants to thank the
Deutsche Forschungs\-ge\-mein\-schaft for financial support.
P.L.\ is grateful to the \"Oster\-reichische Akademie der Wissenschaften
for its support with an APART scholarship.
This work was supported in parts by
the US Department of Energy,  grant no.~DE-FG02-88ER40456.

\newpage


\newpage

\section*{Figure Captions}

{\noindent\bf Figure 1:} One possible tree configuration with random
branching. At each cascade step all existing branches, which have
evolved so far, might independently split into a left and a right
subbranch with probability $\tilde{p}$ or might not split with a
probability $1-\tilde{p}$.

\bigskip

{\noindent\bf Figure 2:} Energy curdling for a binary cascade with
random
branching shown for three cascade steps. After the last cascade step
$(J=3)$ each energy density is resolved at the finest scale.

\bigskip

{\noindent\bf Figure 3:} Reduced two-bin correlation density
$r_{k_1k_2} = \rho_{k_1k_2} / \rho_{k_1} \rho_{k_2}$
of the $p$-model
(a), $\alpha$-model (b), $p$-model with random branching (c) and the
QCD-motivated cascade model (d). The
splitting parameters and the number of cascade steps have been chosen to
be $\alpha=\beta=\gamma_0=0.4$ and $J=6$, respectively;
for case (c) the branching probability has been set to $\tilde{p}=0.5$.

\bigskip

{\noindent\bf Figure 4:} Deviations from scaling of the backward
moments $M_2(J,j)$ (upper two
curves) and Haar wavelet correlations $\tilde{\rho}_{(j,k),(j,k)}^{(J)}$
(lower two curves) with respect to the scale index $j$ for the $p$-
(solid line) and $\alpha$-model (dashed line).
The number of cascade steps is $J=10$ and
$\alpha=\beta=0.4$.

\bigskip

{\noindent\bf Figure 5:} Deviations from scaling of the normalized
forward moments $m_2(j)$ (dashed line), the normalized backward moments
$M_2^{\rm n}(J,j) = M_2(J,j) / (\rho_0^{(J)})^2$ (solid line) and the
normalized Haar wavelet correlations
$\tilde{r}_{(jk),(jk)}^{(J)} =
 \tilde{\rho}_{(jk),(jk)}^{(J)} /
 (\tilde{\rho}_0^{(J)})^2$ (dotted line)
with respect to the scale index $j$ for the QCD-motivated cascade
model. The number of cascade steps is $J=10$ and
$\gamma_0=0.4$.

\bigskip

{\noindent\bf Figure 6:} Reduced Haar wavelet transformed two-bin
correlation density
$\tilde{r}_{(j_1k_1),(j_2k_2)}^{(J)} =
 \tilde{\rho}_{(j_1k_1),(j_2k_2)}^{(J)} /
 (\tilde{\rho}_0^{(J)})^2$
of the $p$-model
(a), $\alpha$-model (b), $p$-model with random branching (c) and the
QCD-motivated cascade model (d). The
splitting parameters and the number of cascade steps have been chosen to
be $\alpha=\beta=\gamma_0=0.4$ and $J=6$, respectively; for case (c) the
branching probability has been set to $\tilde{p}=0.5$.

\bigskip

{\noindent\bf Figure 7:} Deviations from scaling of
$M_2(J,j)$ (upper curves) and
$\tilde{\rho}_{(j,k)(j,k)}^{(J)}$ (lower curves)
for the $p$-model with random branching with $J=10$, $\alpha=0.4$,
$\tilde{p}=1$ (solid curve),
$\tilde{p}=1/(1+\alpha^2)$ (dashed curve) and
$\tilde{p}=0.5$ (dotted curve).

\bigskip

{\noindent\bf Figure 8:} The $p$-model evolution of the energy
densities at three adjacent cascade steps. The marked arrows represent
the difference in energy density of neighbouring bins and are equal to
the corresponding Haar wavelet amplitudes $2 \tilde{\epsilon}_{jk}$.
The clump correlations relate the fluctuations of e.g.\ arrow 1 with
those of arrow 2.

\bigskip

{\noindent\bf Figure 9:} Multiresolution analysis with the Haar wavelet
basis of one particular two-dimensional $\alpha$-model realization at
scale $J=6$ down to $j=2$. Left column:
sequence of smoothing operations. Middle column: difference between two
adjacent smoothed scales. Right column: clump structure of middle column
emphasized by reduction to two gray values (white/black) for
positive/negative regions in the difference information.

\bigskip

{\noindent\bf Figure 10:} Multiresolution analysis of 500 points from a
Poisson transformed $\alpha$-model realization in two dimensions with
respect to the smooth compact Daubechies D12 wavelet. For further
details see Fig.\ 9.


\begin{thebibliography}{99}

\bibitem{MAN74}
  B.\ Mandelbrot:
  J.\ Fluid Mech.\ 62 (1974) 719

\bibitem{FRI78}
  U.\ Frisch, P.L.\ Sulem and M.\ Nelkin:
  J.~Fluid.~Mech.\ 87 (1978)~719

\bibitem{SCH85}
  D.\ Schertzer and S.\ Lovejoy:
  in ``Turbulent Shear Flows 4",
  eds.\ L.J.S.\ Bradbury, F.\ Durst, B.\ Launder, F.W.\ Schmidt and
  J.H.\ Whitelaw,
  pp.\ 7-33, Springer 1985

\bibitem{MEN87}
  C.\ Meneveau and K.R.\ Sreenivasan:
  Phys.~Rev.~Lett.\ 59 (1987)~1424

\bibitem{BIA86}
  A.\ Bia\l as and R.\ Peschanski:
  Nucl.~Phys.\ B273 (1986)~703;
  Nucl.~Phys.\ B308 (1988)~857

\bibitem{LIP89}
  P.\ Lipa and B.\ Buschbeck:
  Phys.~Lett.\ B223 (1989)~465;
  F.\ Botterweck:
  PhD thesis, University Nijmegen 1992;
  M.\ Charlet:
  PhD thesis, University Nijmegen 1994

\bibitem{OCH90}
  W.\ Ochs:
  Phys.\ Lett.\ B247 (1990) 101

\bibitem{BIY90}
  M.\ Biyajima, A.\ Bartl, T.\ Mizoguchi and N.\ Suzuki:
  Phys.\ Lett.\ B237 (1990) 563;
  Phys.\ Lett.\ B247 (1990) 629

\bibitem{FED88}
  J.\ Feder:
  Fractals,
  New York: Plenum Press 1988

\bibitem{MAL89}
  S.\ Mallat:
  IEEE Trans.\ Pattern Anal.\ and Machine Intell.\ 11 (1989)~674

\bibitem{DAU88}
  I.\ Daubechies:
  Comm.\ Pure Appl.\ Math.\ 41 (1988)~909;
  Ten Lectures on Wavelets,
  Philadelphia:
  Society for Industrial and Applied Mathematics (SIAM) 1992

\bibitem{MEY92}
  Y.~Meyer:
  Wavelets and Operators, New York: Cambridge University Press 1992;
  Wavelets: Algorithms and Applications, Philadelphia:
  Society for Industrial and Applied Mathematics (SIAM) 1993

 \bibitem{RUS92}
  M.B.\ Ruskai, G.\ Beylkin, R.\ Coifman, I.\ Daubechies, S.\ Mallat,
  Y.\ Meyer and L.\ Raphael:
  Wavelets and Their Application,
  Boston: Jones and Bartlett 1992

 \bibitem{KAI94}
  G.\ Kaiser:
  A Friendly Guide to Wavelets,
  Boston: Birkh\"auser 1994

\bibitem{FLA92}
  P.~Flandrin:
  IEEE Trans.~Inf.~Theory  38 (1992) 910

\bibitem{GRE94}
  M.~Greiner, P.~Lipa and P.~Carruthers:
  Phys.~Rev.~E, in press

\bibitem{BRA94}
  P.~Brax, J.~Meunier and R.~Peschanski:
  Z.\ Phys.\ C62 (1994) 649;
  J.\ Meunier:
  in Proceedings of the Cracow Workshop on Multiparticle Production
  ``Soft Physics and Fluctuations",
  eds.\ A.\ Bia\l as, K.\ Fialkowski, K.\ Zalewski and R.\ Hwa,
  Singapore: World Scientific 1994, p.\ 261

\bibitem{MEU92}
  J.\ Meunier and R.\ Peschanski:
  Nucl.\ Phys.\ B374 (1992) 327;
  Y.\ Gabellini, J.\ Meunier and R.\ Peschanski:
  Z.\ Phys.\ C55 (1992) 455

\bibitem{CVI80}
  P.~Cvitanovi\'c, P.~Hoyer and K.~Zalewski:
  Nucl.\ Phys.\ B176 (1980) 429

\bibitem{CHU90}
  C.\ Chui and R.~Hwa:
  Phys.\ Lett.\ B236 (1990) 466

\bibitem{OCH92}
  W.\ Ochs and J.\ Wosieck:
  Phys.~Lett.\ 289 (1992)~159;
  304 (1993)~144;
  preprint MPI-PhT/94-94

\bibitem{GUS91}
  G.\ Gustafsson and A.\ Nillson:
  Z.\ Phys.\ C52 (1991) 533

\bibitem{DOK93}
  Y.\ Dokshitzer and I.\ Dremin:
  Nucl.\ Phys.\ B402 (1993) 139

\bibitem{FIE89}
  R.\ Field:
  Applications of Perturbative QCD,
  Redwood City: Addison-Wesley 1989

\bibitem{MAL91}
  S.\ Mallat:
  IEEE Trans.~Inf.~Theory 37 (1991)~1019

\bibitem{MAL92}
  S.\ Mallat and S.\ Zhong:
  IEEE Trans.~Patt.~Anal.~Mach.~Intel.\ 14 (1992)~710

\bibitem{EGG93}
  H.C.\ Eggers, P.\ Lipa, P.\ Carruthers and B.\ Buschbeck:
  Phys.\ Rev.\ D48 (1993) 2040

\bibitem{KAN95}
  J.\ Kantelhardt, M.\ Greiner and E.\ Roman:
  in preparation


\end{thebibliography}
\end{document}